\documentclass[10pt]{article}
\usepackage{graphicx}
\usepackage{amsmath}
\usepackage{amssymb}
\usepackage{caption2}
\begin{document}
\begin{center}
{\large {\bf \sc{ The $Z_b(10610)$  and $Z_b(10650)$ as   axial-vector tetraquark states
in the QCD sum rules }}} \\[2mm]
Zhi-Gang  Wang$^{1}$ \footnote{E-mail: zgwang@aliyun.com.  }, Tao Huang$^{2}$ \footnote{Email: huangtao@ihep.ac.cn}     \\
$^{1}$ Department of Physics, North China Electric Power University, Baoding 071003, P. R. China \\
$^{2}$ Institute of High Energy Physics and Theoretical Physics
Center for Science Facilities, Chinese Academy of Sciences, Beijing 100049, P.R. China
\end{center}

\begin{abstract}
In this article, we study the axial-vector mesons $Z_b(10610)$ and $Z_b(10650)$ with the $C\gamma_\mu-C\gamma_5$ type and $C\gamma_\mu-C\gamma_\nu$ type interpolating currents respectively by carrying out the operator product expansion to  the vacuum condensates up to dimension-10. In calculations,
we explore the energy scale dependence of the QCD spectral densities of the hidden bottom tetraquark states in details for the first time, and suggest a formula $\mu=\sqrt{M^2_{X/Y/Z}-(2{\mathbb{M}}_b)^2}$  with the effective mass ${\mathbb{M}}_b=5.13\,\rm{GeV}$ to determine the energy scales. The numerical results favor assigning the $Z_b(10610)$ and $Z_b(10650)$ as the $C\gamma_\mu-C\gamma_5$ type and $C\gamma_\mu-C\gamma_\nu$ type hidden bottom tetraquark states, respectively. We obtain the mass of the  $J^{PC}=1^{++}$   hidden bottom tetraquark state as a byproduct, which can be compared to the experimental data in the futures.
Furthermore, we  study the strong decays $Z_b^\pm(10610)\to\Upsilon\pi^{\pm}\, ,\,\eta_b\rho^{\pm}$ with the three-point QCD sum rules, the decay widths also support assigning the $Z_b(10610)$ as the $C\gamma_\mu-C\gamma_5$ type hidden bottom tetraquark state.
\end{abstract}



\section{Introduction}

In 2011, the Belle collaboration reported the first observation of the $Z_b(10610)$ and $Z_b(10650)$ in the $\pi^{\pm}\Upsilon({\rm 1,2,3S})$  and $\pi^{\pm} h_b({\rm 1,2P})$  invariant mass distributions  that were produced in association with a single charged pion in $\Upsilon({\rm 5S})$ decays \cite{Belle1105}. The measured masses and widths  are $M_{Z_b(10610)}=\left(10608.4\pm2.0\right)\,\rm{MeV}$, $M_{Z_b(10650)}=\left(10653.2\pm1.5\right)\,\rm{MeV}$, $\Gamma_{Z_b(10610)}=\left(15.6\pm2.5\right)\,\rm{MeV}$ and $\Gamma_{Z_b(10650)}=\left(14.4\pm3.2\right)\,\rm{MeV}$, respectively. The quantum numbers   $I^G(J^P)=1^+(1^+)$ are favored \cite{Belle1105}.
Later, the Belle collaboration updated the measured parameters  $ M_{Z_b(10610)}=(10607.2\pm2.0)\,\rm{ MeV}$, $M_{Z_b(10650)}=(10652.2\pm1.5)\,\rm{MeV}$, $\Gamma_{Z_b(10610)}=(18.4\pm2.4) \,\rm{MeV}$ and
$\Gamma_{Z_b(10650)}=(11.5\pm2.2)\,\rm{ MeV}$ \cite{Belle1110}.
In 2013, the Belle collaboration observed the $\Upsilon(5{\rm S}) \to \Upsilon ({\rm 1,2,3S}) \pi^0 \pi^0$ decays for the first time, and obtained  the neutral partner of the $Z_b^{\pm}(10610)$, the $Z_b^0(10610)$, in a Dalitz
analysis of the decays to $\Upsilon(2,3{\rm S}) \pi^0$ \cite{Belle1308}.
There have been several tentative assignments of the  $Z_b(10610)$ and $Z_b(10650)$, such as the molecular states \cite{Molecule-Zb}, tetraquark states \cite{Tetraquark-Zb,Tetraquark-Zb-QCDSR}, threshold cusps \cite{Cusp-Zb}, the re-scattering effects \cite{Rescatter-Zb}, etc.

In 2013, the BESIII collaboration observed the $Z_c^{\pm}(3900)$ in the $\pi^\pm J/\psi$ mass spectrum in the process  $e^+e^- \to \pi^+\pi^-J/\psi$ \cite{BES3900},  then the $Z_c^{\pm}(3900)$ was confirmed by the Belle and CLEO collaborations \cite{Belle3900,CLEO3900}.
   Later, the BESIII collaboration  observed
the $Z^{\pm}_c(4025)$ near the $(D^{*} \bar{D}^{*})^{\pm}$ threshold in the $\pi^\mp$ recoil mass spectrum  in the process $e^+e^- \to (D^{*} \bar{D}^{*})^{\pm} \pi^\mp$ \cite{BES1308}.
Furthermore, the  BESIII collaboration observed the  $Z_c^{\pm}(4020)$   in the $\pi^\pm h_c$ mass spectrum in the process $e^+e^- \to \pi^+\pi^- h_c$  \cite{BES1309}.
The $Z_b(10610)$, $Z_b(10650)$, $Z_c(3900)$ and $Z_c(4020)$ are observed in the analogous decays to the final states $\pi^{\pm}\Upsilon({\rm 1,2,3S})$, $\pi^{\pm} h_b({\rm 1,2P})$, $\pi^\pm J/\psi$, $\pi^\pm h_c$, and should have analogous structures.

In Refs.\cite{WangHuangTao,Wang1311,Wang1312}, we distinguish
the charge conjugations of the interpolating  currents, calculate the  vacuum condensates up to dimension-10  in the operator product expansion, study the diquark-antidiquark type scalar, vector, axial-vector and tensor hidden charmed tetraquark states in a systematic way  with the QCD sum rules, make reasonable  assignments of the $X(3872)$, $Z_c(3900)$, $Z_c(3885)$, $Z_c(4020)$, $Z_c(4025)$, $Z(4050)$, $Z(4250)$, $Y(4360)$, $Y(4630)$ and $Y(4660)$.  Furthermore, we    explore the energy scale dependence of the hidden charmed tetraquark states  in details for the first time, and suggest a  formula,
\begin{eqnarray}
\mu&=&\sqrt{M^2_{X/Y/Z}-(2{\mathbb{M}}_c)^2} \, ,
 \end{eqnarray}
 with the effective mass ${\mathbb{M}}_c=1.8\,\rm{GeV}$ to determine the energy scales of the  QCD spectral densities.
 The numerical results  favor assigning   the $X(3872)$ and $Z_c(3900)$ (or $Z_c(3885)$) as the $1^{++}$ and $1^{+-}$ diquark-antidiquark type  tetraquark states, respectively,  and assigning  the $Z_c(4020)$ and $Z_c(4025)$   as the $J^{PC}=1^{+-}$ or  $2^{++}$   diquark-antidiquark type tetraquark states.

The diquarks have  five Dirac tensor structures, scalar $C\gamma_5$,
pseudoscalar $C$, vector $C\gamma_\mu \gamma_5$, axial vector
$C\gamma_\mu $  and  tensor $C\sigma_{\mu\nu}$. In Ref.\cite{Wang-Axial}, we study the $C\gamma_5-C\gamma_\mu$ type axial-vector hidden charmed and hidden bottom tetraquark states with the QCD sum rules,  obtain the ground state mass    $M_{b\bar{b}u\bar{d}}=(11.27 \pm 0.20)\,\rm{GeV}$, where the charge conjugations are not distinguished, the $\overline{MS}$ quark mass  $\overline{m}_b(\mu={1\rm GeV}) = (4.8 \pm 0.1)\,\rm{GeV}$ is chosen. The energy scale $\mu=1\,\rm{GeV}$ is somewhat too small. The predictions $M_{b\bar{b}u\bar{d}}-M_{Z_b(10610)}= (0.66\pm0.20)\,\rm{GeV}$ and $M_{b\bar{b}u\bar{d}}-M_{Z_b(10650)}= (0.62\pm0.20)\,\rm{GeV}$ disfavor  assigning the $Z_b(10610)$ and $Z_b(10650)$  as the axial-vector tetraquark states.
In Ref.\cite{Tetraquark-Zb-QCDSR}, Cui, Liu and Huang distinguish the charge conjugations, study the $C\gamma_5-C\gamma_\mu$ and $\epsilon^{\mu\nu\alpha\beta}\left(C\gamma_\nu -\partial_\alpha-C\gamma_\beta\right)$ type axial-vector hidden  bottom tetraquark states with the QCD sum rules by carrying out the operator product expansion up to the vacuum condensates of dimension 6. Their predictions  favor  assigning the $Z_b(10610)$ and $Z_b(10650)$  as the axial-vector tetraquark states. However, the  energy scales of the QCD spectral densities are not shown or not specified \cite{Tetraquark-Zb-QCDSR}. In Ref.\cite{Tetraquark-Zb-QCDSR} (\cite{Wang-Axial})  higher (some higher) dimension vacuum condensates are neglected. There appear terms of the orders $\mathcal{O}\left(\frac{1}{T^2}\right)$, $\mathcal{O}\left(\frac{1}{T^4}\right)$, $\mathcal{O}\left(\frac{1}{T^6}\right)$ in the QCD spectral densities, if we take into account the vacuum condensates whose dimensions are larger than 6.  The terms associate with $\frac{1}{T^2}$, $\frac{1}{T^4}$, $\frac{1}{T^6}$ in the QCD spectral densities  manifest themselves at small values of the Borel parameter $T^2$, we have to choose large values of the $T^2$ to warrant convergence of the operator product expansion and appearance of the Borel platforms. In the Borel windows, the higher dimension vacuum condensates  play a less important role.
In summary, the higher dimension vacuum condensates play an important role in determining the Borel windows therefore the ground state  masses and pole residues, so we should take them into account consistently.

 In this article, we extend our previous works in Refs.\cite{WangHuangTao,Wang1311,Wang1312} to study the $C\gamma_\mu-C\gamma_5$ type and $C\gamma_\mu-C\gamma_\nu$ type  axial-vector tetraquark states by calculating the vacuum condensates up to dimension-10 in a systematic way,   make reasonable  assignments of the $Z_b(10610)$  and $Z_b(10650)$  based on the QCD sum rules. Furthermore, we extend  the energy scale formula to study the hidden bottom diquark-antidiquark systems,
 \begin{eqnarray}
\mu&=&\sqrt{M^2_{X/Y/Z}-(2{\mathbb{M}}_b)^2} \, ,
 \end{eqnarray}
 and   make  efforts to explore the energy scale dependence in details for the first time, and try to fit the effective mass ${\mathbb{M}}_b$.

The article is arranged as follows:  we derive the QCD sum rules for
the masses and pole residues of  the axial-vector tetraquark states  in section 2;
in section 3, we present the numerical results and discussions; in section 4, we study the strong decays $Z_b^\pm(10610)\to\Upsilon\pi^{\pm}, \, \, \eta_b\rho^{\pm}$ with the three-point QCD sum rules; section 5 is reserved for our conclusion.

\section{QCD sum rules for  the   $J^{PC}=1^{+\pm}$ tetraquark states }
In the following, we write down  the two-point correlation functions $\Pi_{\mu\nu}(p)$ and $\Pi_{\mu\nu\alpha\beta}(p)$ in the QCD sum rules,
\begin{eqnarray}
\Pi_{\mu\nu}(p)&=&i\int d^4x e^{ip \cdot x} \langle0|T\left\{J_\mu(x)J_\nu^{\dagger}(0)\right\}|0\rangle \, , \\
\Pi_{\mu\nu\alpha\beta}(p)&=&i\int d^4x e^{ip \cdot x} \langle0|T\left\{J_{\mu\nu}(x)J_{\alpha\beta}^{\dagger}(0)\right\}|0\rangle \, ,
\end{eqnarray}
\begin{eqnarray}
J_\mu(x)&=&\frac{\epsilon^{ijk}\epsilon^{imn}}{\sqrt{2}}\left\{u^j(x)C\gamma_5b^k(x) \bar{d}^m(x)\gamma_\mu C \bar{b}^n(x)+tu^j(x)C\gamma_\mu b^k(x)\bar{d}^m(x)\gamma_5C \bar{b}^n(x) \right\} \, ,\\
J_{\mu\nu}(x)&=&\frac{\epsilon^{ijk}\epsilon^{imn}}{\sqrt{2}}\left\{u^j(x)C\gamma_\mu b^k(x) \bar{d}^m(x)\gamma_\nu C \bar{b}^n(x)-u^j(x)C\gamma_\nu b^k(x)\bar{d}^m(x)\gamma_\mu C \bar{b}^n(x) \right\} \, ,
\end{eqnarray}
 the $i$, $j$, $k$, $m$, $n$ are color indexes, and the $C$ is the charge conjugation matrix.
Under charge conjugation transform $\widehat{C}$, the currents $J_\mu(x)$ and $J_{\mu\nu}(x)$ have the properties,
\begin{eqnarray}
\widehat{C}J_{\mu}(x)\widehat{C}^{-1}&=&\pm J_{\mu}(x)\mid_{u\leftrightarrow  d } \,\,\,\, {\rm for}\,\,\,\, t=\pm1\, , \nonumber\\
\widehat{C}J_{\mu\nu}(x)\widehat{C}^{-1}&=&- J_{\mu\nu}(x)\mid_{u\leftrightarrow  d } \, ,
\end{eqnarray}
$t=\pm 1$ correspond to the positive and negative charge conjugations, respectively.
We choose  the $C\gamma_\mu-C\gamma_5$ type (type I) currents $J_\mu(x)$ to interpolate the
     tetraquark state $Z_b(10610)$ with $J^{PC}=1^{+-}$ and its charge conjugation partner with $J^{PC}=1^{++}$.  Furthermore, we  choose  the $C\gamma_\mu-C\gamma_\nu$ type (type II) current $J_{\mu\nu}(x)$ to interpolate the     tetraquark state  $Z_b(10650)$ with $J^{PC}=1^{+-}$. In Refs.\cite{WangHuangTao,Wang1312}, we observe that
   the type II axial-vector hidden-charmed tetraquark states have larger masses than that of the  type I. We expect that the type II axial-vector hidden-bottom tetraquark states also have larger masses than that of the  type I. There are other routines to construct the axial-vector currents \cite{Chenwei}.

We can insert  a complete set of intermediate hadronic states with
the same quantum numbers as the current operators $J_\mu(x)$ and $J_{\mu\nu}(x)$ into the
correlation functions $\Pi_{\mu\nu}(p)$ and $\Pi_{\mu\nu\alpha\beta}(p)$  to obtain the hadronic representation
\cite{SVZ79,Reinders85}. After isolating the ground state
contributions from the axial-vector (and vector) tetraquark states, we get the following results,
\begin{eqnarray}
\Pi_{\mu\nu}(p)&=&\Pi^{\rm I}(p)\left(-g_{\mu\nu}+\frac{p_\mu p_\nu}{p^2}\right) +\Pi_0(p)\frac{p_\mu p_\nu}{p^2} \, \, , \nonumber\\ &=&\frac{\lambda_Z^2}{M_Z^2-p^2}\,\left(-g_{\mu\nu}+\frac{p_\mu p_\nu}{p^2}\right)   +\cdots \, \, ,\\
\Pi_{\mu\nu\alpha\beta}(p)&=&\Pi^{\rm II}(p)\left( -g_{\mu\alpha}p_{\nu}p_{\beta}-g_{\nu\beta}p_{\mu}p_{\alpha}+g_{\mu\beta}p_{\nu}p_{\alpha}+g_{\nu\alpha}p_{\mu}p_{\beta}\right) +\nonumber\\
&&\Pi_{-}(p)\left(p^2g_{\mu\alpha}g_{\nu\beta}-p^2g_{\mu\beta}g_{\nu\alpha} -g_{\mu\alpha}p_{\nu}p_{\beta}-g_{\nu\beta}p_{\mu}p_{\alpha}+g_{\mu\beta}p_{\nu}p_{\alpha}+g_{\nu\alpha}p_{\mu}p_{\beta}\right) \, \, , \nonumber\\
&=&\frac{\lambda_{ Z}^2}{M_{Z}^2-p^2}\left( -g_{\mu\alpha}p_{\nu}p_{\beta}-g_{\nu\beta}p_{\mu}p_{\alpha}+g_{\mu\beta}p_{\nu}p_{\alpha}+g_{\nu\alpha}p_{\mu}p_{\beta}\right) + \nonumber\\
&&\frac{\lambda_{ Z^\prime}^2}{M_{Z^\prime}^2-p^2}\left(p^2g_{\mu\alpha}g_{\nu\beta}-p^2g_{\mu\beta}g_{\nu\alpha} -g_{\mu\alpha}p_{\nu}p_{\beta}-g_{\nu\beta}p_{\mu}p_{\alpha}+g_{\mu\beta}p_{\nu}p_{\alpha}+g_{\nu\alpha}p_{\mu}p_{\beta}\right)\nonumber\\
&&+\cdots \, \, ,
\end{eqnarray}
where the spin-0 component   $\Pi_0(p)$ and the spin-1 component $\Pi_{-}(p)$ are irrelevant in the present analysis \cite{WangHcHb}, the pole residues   $\lambda_Z$ ($\lambda_{Z^\prime}$) are defined by
\begin{eqnarray}
 \langle 0|J_\mu(0)|Z(p)\rangle&=&\lambda_{Z} \, \varepsilon_\mu \, ,\nonumber\\
 \langle 0|J_{\mu\nu}(0)|Z(p)\rangle &=& \lambda_{Z}\left(\varepsilon_{\mu}p_{\nu}-\varepsilon_{\nu}p_{\mu} \right)\, ,\nonumber\\
 \langle 0|J_{\mu\nu}(0)|Z^{\prime}(p)\rangle &=& \lambda_{Z^\prime}\epsilon_{\mu\nu\alpha\beta}\varepsilon^{\alpha}p^{\beta}\, ,
\end{eqnarray}
the $\varepsilon_\mu$ are the polarization vectors of the axial-vector (and vector) tetraquark states. The current $J_{\mu\nu}$ has  non-vanishing couplings both to the
$J^{PC}=1^{+-}$ tetraquark state $Z$ and the $J^{PC}=1^{--}$ tetraquark state $Z^{\prime}$. In Refs.\cite{Wang1311,Wang1312}, we observe that the energy gaps between the vector and axial-vector hidden charmed tetraquark states are about $0.65\,\rm{GeV}$ based on the QCD  sum rules. So we expect that the energy gaps between the vector and axial-vector hidden bottom tetraquark states are also about $0.65\,\rm{GeV}$, the vector tetraquark state $Z^{\prime}$ has no contamination.

The current-meson (or baryon) duality is a basic assumption of the QCD sum rules, the current couples potentially to a special hadron. The two-point QCD sum rules can neither prove nor disprove the existence of  the special hadron strictly, but can give reasonable mass and pole residue to be confronted with the experimental data.
Furthermore, we can take the pole residue as basic input parameter to study the relevant processes with the three-point QCD sum rules, the predictions can also be confronted with the experimental data and shed light on the nature of the special hadron. In the present case, the predicted masses maybe favor or disfavor assigning the $Z_b(10610)$ and $Z_b(10650)$ as the axial-vector tetraquark states, while the predicted hadronic coupling constants therefore the decay widths serve as additional constraints in assigning the $Z_b(10610)$ and $Z_b(10650)$.

 We carry out the operator product expansion up to the vacuum condensates of  dimension-10, then obtain the QCD spectral densities through dispersion relation,
 take the quark-hadron duality below the thresholds $s_0$, and perform Borel transform  with respect to
the variable $P^2=-p^2$ to obtain  the  QCD sum rules:
\begin{eqnarray}
\lambda^2_Z\, e^{-\frac{M^2_Z}{T^2}}= \int_{4m_b^2}^{s_0} ds\, \rho(s) \, e^{-\frac{s}{T^2}} \, ,
\end{eqnarray}
where
\begin{eqnarray}
\rho(s)&=&\rho_{0}(s)+\rho_{3}(s) +\rho_{4}(s)+\rho_{5}(s)+\rho_{6}(s)+\rho_{7}(s) +\rho_{8}(s)+\rho_{10}(s)\, ,
\end{eqnarray}
the subscripts  $0$, $3$, $4$, $5$, $6$, $7$, $8$, $10$ denote the dimensions of the  vacuum condensates, the explicit expressions are presented in the Appendix. One can consult Refs.\cite{WangHuangTao,Wang1312} for the technical details.

 Differentiate   Eq.(11) with respect to  $\frac{1}{T^2}$, then eliminate the
 pole residues $\lambda_Z$, we obtain the QCD sum rules for
 the masses of the axial-vector hidden bottom tetraquark states,
 \begin{eqnarray}
 M_Z^2= \frac{\int_{4m_b^2}^{s_0} ds\frac{d}{d \left(-1/T^2\right)}\rho(s)e^{-\frac{s}{T^2}}}{\int_{4m_b^2}^{s_0} ds \rho(s)e^{-\frac{s}{T^2}}}\, .
\end{eqnarray}

\section{Numerical results and discussions}
In this article, we study the energy scale dependence of the QCD spectral densities of the hidden bottom tetraquark states in details for the first time and search for the ideal energy scales $\mu$ of the QCD spectral densities.

The initial input parameters are taken to be the standard values $\langle
\bar{q}q \rangle=-(0.24\pm 0.01\, \rm{GeV})^3$,   $\langle
\bar{q}g_s\sigma G q \rangle=m_0^2\langle \bar{q}q \rangle$,
$m_0^2=(0.8 \pm 0.1)\,\rm{GeV}^2$, $\langle \frac{\alpha_s
GG}{\pi}\rangle=(0.33\,\rm{GeV})^4 $    at the energy scale  $\mu=1\, \rm{GeV}$ from the QCD sum rules
\cite{SVZ79,Reinders85,Ioffe2005,ColangeloReview}, and $m_{b}(m_b)=(4.18\pm0.03)\,\rm{GeV}$
 from the Particle Data Group \cite{PDG}.  We take into account
the energy-scale dependence of  the quark condensate, mixed quark condensate and $\overline{MS}$ mass from the renormalization group equation,
 \begin{eqnarray}
 \langle\bar{q}q \rangle(\mu)&=&\langle\bar{q}q \rangle(Q)\left[\frac{\alpha_{s}(Q)}{\alpha_{s}(\mu)}\right]^{\frac{4}{9}}\, ,\nonumber\\
\langle\bar{q}g_s \sigma Gq \rangle(\mu)&=&\langle\bar{q}g_s \sigma Gq \rangle(Q)\left[\frac{\alpha_{s}(Q)}{\alpha_{s}(\mu)}\right]^{\frac{2}{27}} \, ,\nonumber\\
m_b(\mu)&=&m_b(m_b)\left[\frac{\alpha_{s}(\mu)}{\alpha_{s}(m_b)}\right]^{\frac{12}{23}} \, ,\nonumber\\
\alpha_s(\mu)&=&\frac{1}{b_0t}\left[1-\frac{b_1}{b_0^2}\frac{\log t}{t} +\frac{b_1^2(\log^2{t}-\log{t}-1)+b_0b_2}{b_0^4t^2}\right]\, ,
\end{eqnarray}
  where $t=\log \frac{\mu^2}{\Lambda^2}$, $b_0=\frac{33-2n_f}{12\pi}$, $b_1=\frac{153-19n_f}{24\pi^2}$, $b_2=\frac{2857-\frac{5033}{9}n_f+\frac{325}{27}n_f^2}{128\pi^3}$,  $\Lambda=213\,\rm{MeV}$, $296\,\rm{MeV}$  and  $339\,\rm{MeV}$ for the flavors  $n_f=5$, $4$ and $3$, respectively  \cite{PDG}.

 In QCD, the perturbative quark propagator in the momentum space can be written as
\begin{eqnarray}
S(p)&=&\frac{i}{\!\not\!{p}-m^0-\Sigma(\!\not\!{p},m^0)}\, ,
\end{eqnarray}
where the $m^0$ is the bare mass and the $\Sigma(\!\not\!{p},m^0)$ is the self-energy comes from the one-particle irreducible Feynman diagrams. The renormalized mass $m_r$ is defined as $m^0=m_r+\delta m$. It is convenient to choose the $\overline{MS}$ renormalization scheme by using the counterterm $\delta m$ to absorb the ultraviolet divergences of the form $\left[1/\epsilon+\log4\pi-\gamma_E\right]^L$, $L=1,2,\cdots$, then the $m_r$ is the $\overline{MS}$ mass. On the other hand, we can also define the pole mass by the setting $\!\not\!{p}-m^0-\Sigma(\!\not\!{p},m^0)=0$ with the on-shell mass $\!\not\!{p}=m$. The pole mass and the $\overline{MS}$ mass have the relation $m-m_r=\delta m+\Sigma(m,m^0)$. In QED, the electron mass is a directly observable quantity,  the pole mass is the physical mass and it is more convenient to choose the pole mass. While in QCD, the quark mass is not a directly observable quantity,  we have two choices (choosing $\overline{MS}$ mass or pole mass) in perturbative calculations. However, the pole mass $m_b=(4.78 \pm 0.06)\,\rm{GeV}$ \cite{PDG} leads to much smaller integral range $\int_{4m_b^2}^{s_0}$ of $ds$ in the present case, which does not warrant reasonable QCD sum rules; the pole mass is not preferred.   If the perturbative corrections are neglected, we can also choose other values besides the $\overline{MS}$ mass and pole mass, the mass is just a parameter.

In this article, we neglect the perturbative  ${\mathcal{O}}(\alpha_s)$ corrections to the QCD spectral densities, nevertheless the terms $g_s^2\langle \bar{q}q\rangle^2$ appear; we prefer the $\overline{MS}$ mass. The four-quark condensate $g_s^2\langle \bar{q}q\rangle^2$ comes from the terms
$\langle \bar{q}\gamma_\mu t^a q g_s D_\eta G^a_{\lambda\tau}\rangle$, $\langle\bar{q}_jD^{\dagger}_{\mu}D^{\dagger}_{\nu}D^{\dagger}_{\alpha}q_i\rangle$  and
$\langle\bar{q}_jD_{\mu}D_{\nu}D_{\alpha}q_i\rangle$, rather than comes from the perturbative corrections of $\langle \bar{q}q\rangle^2$ \cite{WangHuangTao}. The $\alpha_s(\mu)=\frac{g_s^2(\mu)}{4\pi}$ is characterized by the energy scale $\mu$, and originates  from the renormalization of the $SU(3)$ color gauge theory.  Furthermore, the condensates $\langle \bar{q}q\rangle$ and $\langle \bar{q}g_s \sigma Gq\rangle$ are scale dependent. It is   convenient to choose the $\overline{MS}$ mass, the QCD spectral densities evolve with  the energy scale $\mu$ consistently.  The present calculations are  directly applicable  when the perturbative corrections are available in the futures.

 In the two-point QCD sum rules for the heavy-light pseudoscalar mesons, neglecting the perturbative ${\mathcal{O}}(\alpha_s)$ corrections to the QCD spectral densities
 can reproduce the  experimental values of the masses but cannot reproduce the experimental values of the decay constants \cite{Wang1301}.
For the tetraquark states, it is more reasonable to refer to the $\lambda_{X/Y/Z}$ as the pole residues (not the decay constants).
We cannot obtain the true values of the pole residues $\lambda_{X/Y/Z}$ by measuring the leptonic decays as in the cases of the $D_{s}(D)$ and $J/\psi (\Upsilon)$,
$D_{s}(D)\to \ell\nu$ and $J/\psi (\Upsilon)\to e^+e^-$, and have to calculate the  $\lambda_{X/Y/Z}$ using  some theoretical methods. It is hard to obtain the true values. In this article, we focus on the masses to study the tetraquark states, and the unknown contributions of the perturbative corrections to the pole residues are canceled out  efficiently when we calculate the hadronic coupling constants (or form-factors) with the three-point QCD sum rules, see Eqs.(34-35). Neglecting perturbative ${\mathcal{O}}(\alpha_s)$ corrections cannot impair the predictive ability qualitatively.

The  bottomonium  states have the masses $M_{\Upsilon}=(9460.30\pm0.26)\,\rm{MeV}$, $M_{\Upsilon^{\prime}}=(10023.26\pm0.31)\,\rm{MeV}$,  $M_{\eta_b}=(9398.0\pm 3.2)\,\rm{MeV}$, $M_{\eta_b^{\prime}}=\left(9999.0 \pm3.5{}^{+2.8}_{-1.9}\right)\,\rm{MeV}$ from the Particle Data Group \cite{PDG}; the energy gaps between the ground states and first radial excited states are about $(0.55-0.60)\,\rm{GeV}$.
In the scenario of tetraquark states,  the  $Z(4430)$ is tentatively assigned to be  the first radial excitation of the $Z_c(3900)$ according to the
analogous decays,
$Z_c(3900)^\pm=J/\psi\pi^\pm$, $Z(4430)^\pm=\psi^\prime\pi^\pm$,
and the mass differences $M_{Z(4430)}-M_{Z_c(3900)}=576\,\rm{MeV}$, $M_{\psi^\prime}-M_{J/\psi}=589\,\rm{MeV}$ \cite{Maiani-2014}; the energy gaps between the ground states and first radial excited states are about $(0.50-0.60)\,\rm{GeV}$. We can estimate that the energy gaps between the ground states and first radial excited states are about $(0.40-0.60)\,\rm{GeV}$ for the hidden bottom tetraquark states based on the heavy quark symmetry.
In this article, we take the threshold parameters as $s_0=(124\pm2)\,\rm{GeV}^2$ and $(125\pm2)\,\rm{GeV}^2$ for the  type I and  type II tetraquark states, respectively, then $\sqrt{s_0}-M_{Z_b(10610)}=(0.4-0.6)\,\rm{GeV}$ and $\sqrt{s_0}-M_{Z_b(10650)}=(0.4-0.6)\,\rm{GeV}$, it is reasonable in the QCD sum rules.  We can also choose  larger continuum threshold parameters, but the contaminations from the higher resonances or continuum states are expected to included in.
On the other  hand, 
the current $J_{\mu\nu}$ has  non-vanishing couplings both to the
$J^{PC}=1^{+-}$ tetraquark state $Z$ and the $J^{PC}=1^{--}$ tetraquark state $Z^{\prime}$, larger continuum threshold parameters maybe result in   contamination from the vector  tetraquark state $Z^{\prime}$.

In Ref.\cite{WangHuangTao,Wang1311,Wang1312}, we study the energy scale dependence of the QCD spectral densities of the hidden charmed tetraquark states  in details for the first time, suggest a formula to estimate the energy scales of the QCD spectral densities in the QCD sum rules,
 $ \mu=\sqrt{M^2_{X/Y/Z}-(2{\mathbb{M}}_c)^2}$, with the effective $c$-quark mass ${\mathbb{M}}_c=1.8\,\rm{GeV}$.
 The heavy tetraquark system could be described
by a double-well potential with two light quarks $q^{\prime}\bar{q}$ lying in the two wells respectively.
   In the heavy quark limit, the $c$ (and $b$) quark can be taken as a static well potential,
which binds the light quark $q^{\prime}$ to form a diquark in the color antitriplet channel or binds the light antiquark $\bar{q}$ to form a meson in the color singlet channel (or a meson-like state in the color octet  channel). Then the heavy tetraquark states  are characterized by the effective heavy quark masses ${\mathbb{M}}_Q$ (or constituent quark masses) and the virtuality $V=\sqrt{M^2_{X/Y/Z}-(2{\mathbb{M}}_Q)^2}$ (or bound energy not as robust). It is natural to take the energy  scale $\mu=V$. The energy scale formula works well for the hidden charmed tetraquark states, we extend the formula to study the energy scales of the QCD spectral densities of the hidden bottom tetraquark states.

\begin{figure}
\centering
\includegraphics[totalheight=6cm,width=7cm]{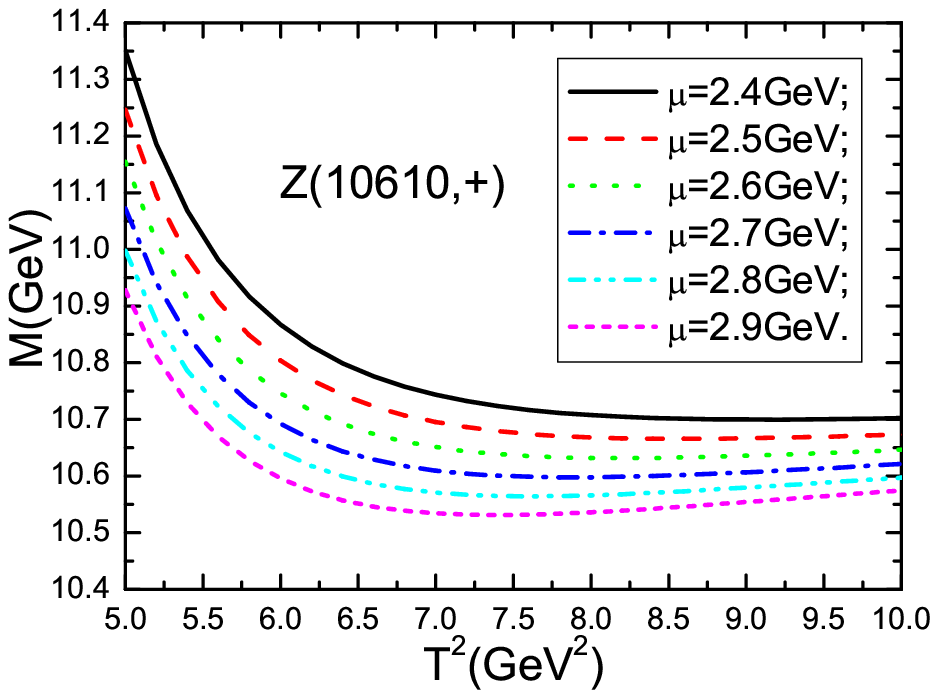}
\includegraphics[totalheight=6cm,width=7cm]{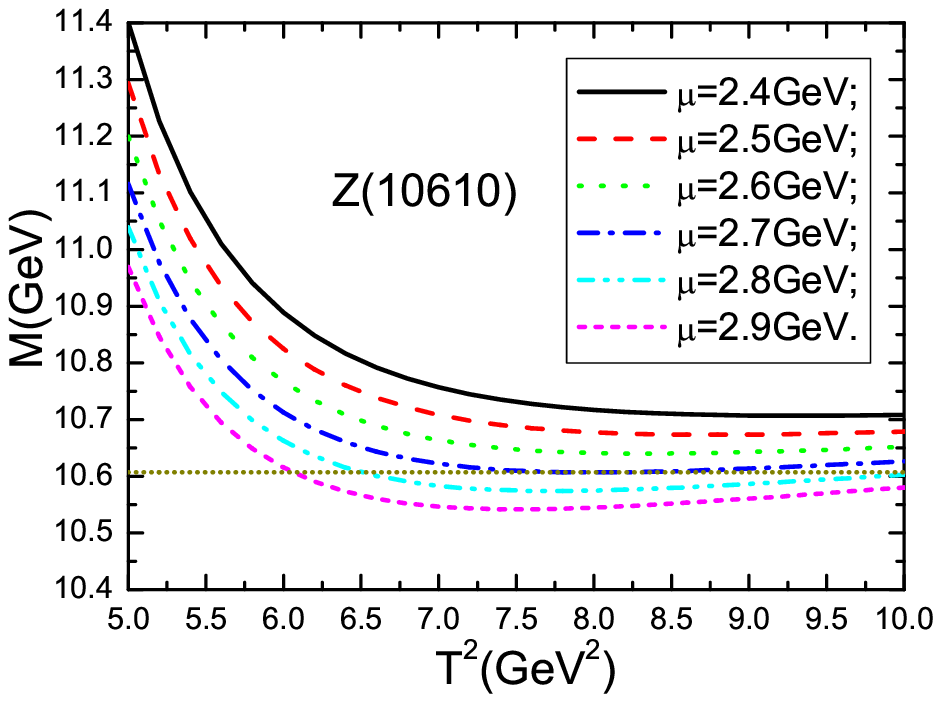}
\includegraphics[totalheight=6cm,width=7cm]{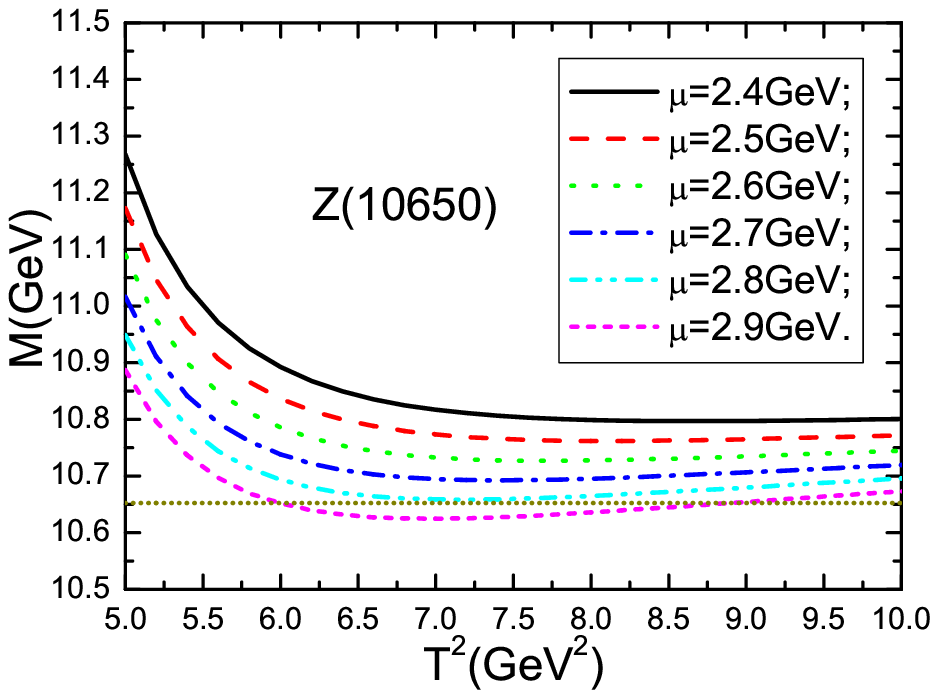}
  \caption{ The masses  with variations of the  Borel parameters $T^2$ and energy scales $\mu$, where the horizontal lines denote the experimental values, the $Z(10610,+)$ denotes the positive charge conjugation partner of the $Z_b(10610)$. }
\end{figure}
In Fig.1,  the masses are plotted   with variations of the  Borel parameters $T^2$ and energy scales $\mu$ for the threshold parameters $s_0=124\,\rm{GeV}^2$ and $s_0=125\,\rm{GeV}^2$ in the cases of the type I and type II tetraquark states, respectively. From the figure, we can see that the masses decrease monotonously with increase of the energy scales, just like that of the hidden charmed tetraquark states \cite{WangHuangTao,Wang1311,Wang1312}. The energy scale $\mu=2.7\,\rm{GeV}$ is the optimal energy scale to reproduce the experimental value  $M_{Z_b(10610)}=10.61\,\rm{GeV}$, then we can fit the parameter ${\mathbb{M}}_b=5.13\,\rm{GeV}$. The resulting  energy scale $\mu=\sqrt{M^2_{Z_b(10650)}-(2\times5.13\,\rm{GeV})^2}=2.85\,\rm{GeV}$ is the optimal energy scale to  reproduce the experimental data $M_{Z_b(10650)}=10.65\,\rm{GeV}$ approximately. The energy scales
$\mu=(2.8-2.9)\,\rm{GeV}$ are the allowed energy scales for the $Z_b(10650)$, see Fig.1; the uncertainty of the energy scale $\mu$ is about $0.05\,\rm{GeV}$. In this article, we take $\delta \mu=0.05\,\rm{GeV}$ for all the hidden bottom tetraquark states.   The energy scale formula $\mu=\sqrt{M^2_{X/Y/Z}-(2{\mathbb{M}}_Q)^2}$ works well, it also works well   for the heavy  molecular states \cite{Wang4140}, the results will be presented elsewhere.

 \begin{figure}
\centering
\includegraphics[totalheight=6cm,width=7cm]{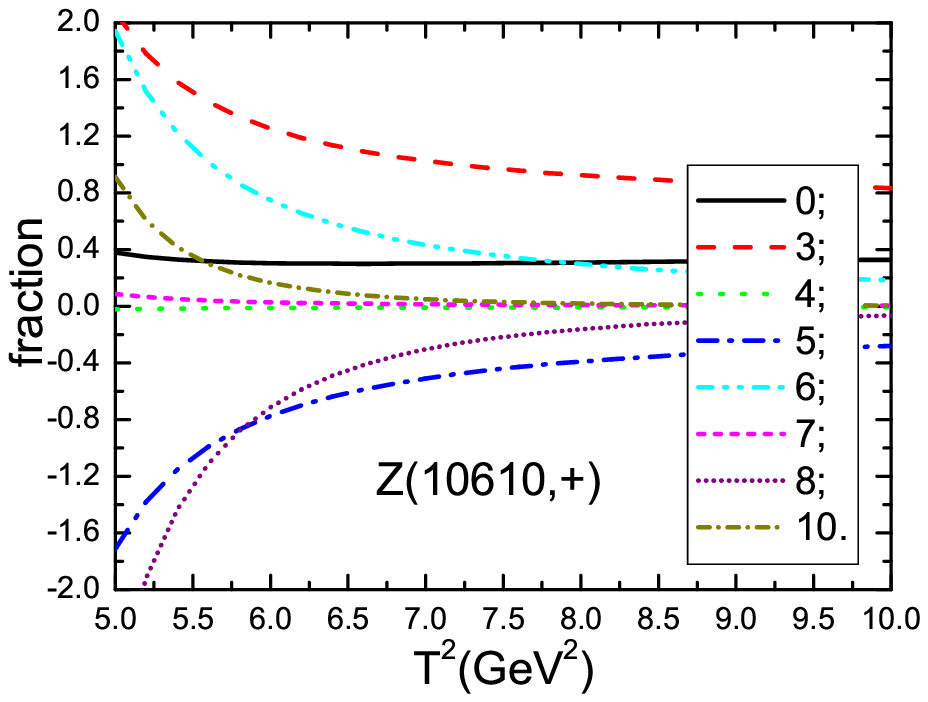}
\includegraphics[totalheight=6cm,width=7cm]{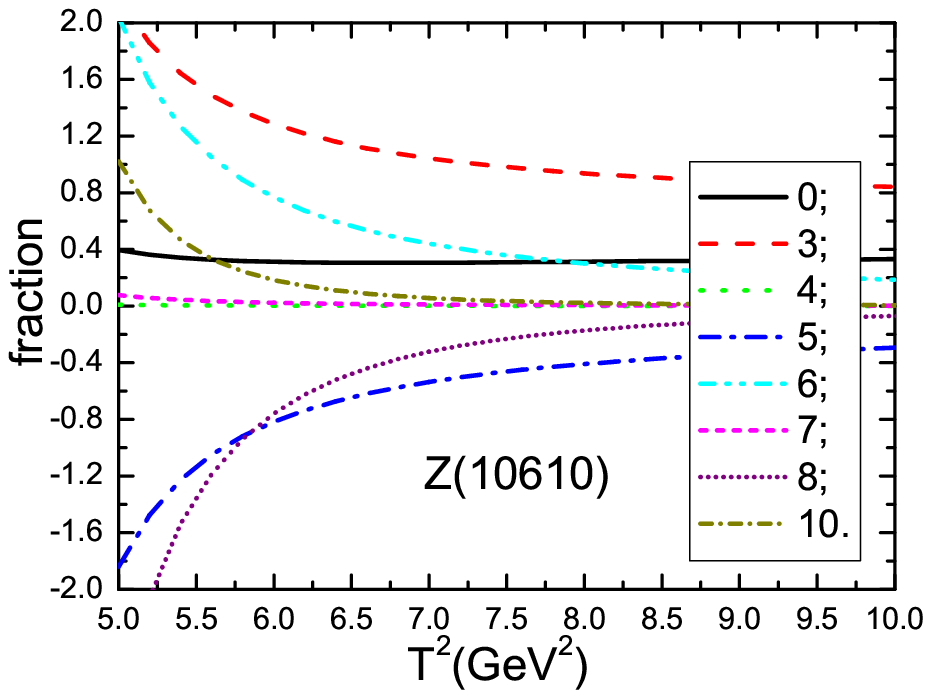}
\includegraphics[totalheight=6cm,width=7cm]{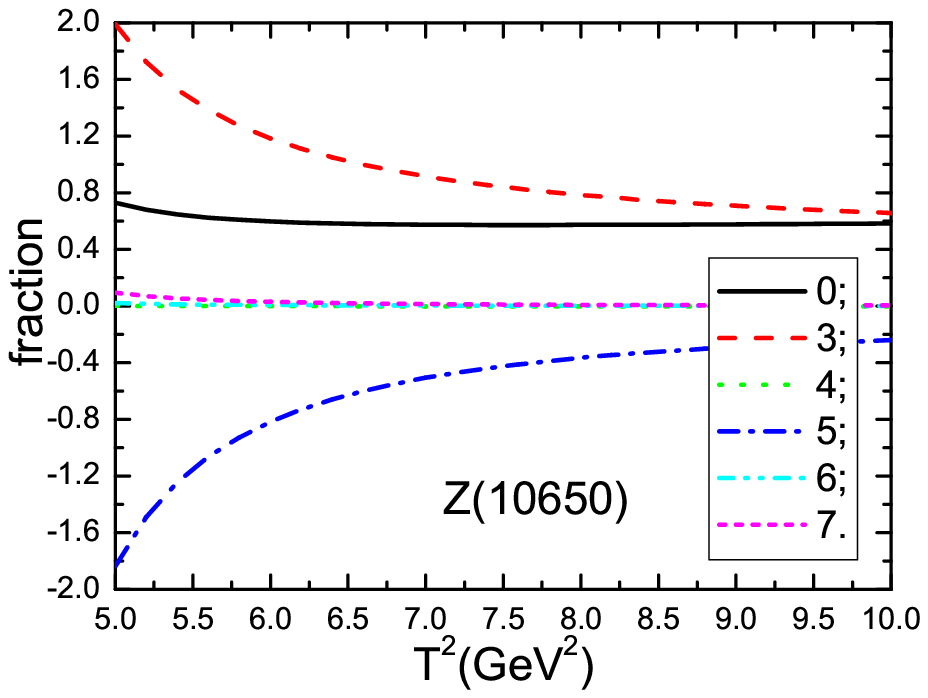}
  \caption{ The contributions of different terms in the operator product expansion  with variations of the  Borel parameters $T^2$, where the 0, 3, 4, 5, 6, 7, 8, 10 denote the dimensions of the vacuum condensates, the $Z(10610,+)$ denotes the positive charge conjugation partner of the $Z_b(10610)$. }
\end{figure}

In Fig.2,  the contributions of different terms in the
operator product expansion are plotted with variations of the Borel parameters  $T^2$ for the parameters $s_0=124\,\rm{GeV}^2$, $\mu=2.70\,\rm{GeV}$ and $s_0=125\,\rm{GeV}^2$, $\mu=2.85\,\rm{GeV}$ in the cases of the type I and type II tetraquark states, respectively. If we take the values  $T^2=(7-8)\,\rm{GeV}^2$, the convergent behavior  is very good.  In Fig.3,  the contributions of the pole terms are plotted with
variations of the threshold parameters $s_0$ and Borel parameters $T^2$ at the energy scales $\mu=2.70\,\rm{GeV}$ and
 $\mu=2.85\,\rm{GeV}$ for the type I and type II tetraquark states, respectively. The values $T^2=(7-8)\,\rm{GeV}^2$ also lead to analogous pole contributions
 $(50-70)\%$. The pole dominance condition is also well satisfied. In Fig.3, the pole contributions are defined by
 \begin{eqnarray}
 {\rm pole} &=& \frac{\int_{4m_b^2}^{s_0}ds\,\rho(s)\exp\left( -\frac{s}{T^2}\right)}{\int_{4m_b^2}^{\infty}ds\,\rho(s)\exp\left( -\frac{s}{T^2}\right)}\, \, .
 \end{eqnarray}

\begin{figure}
\centering
\includegraphics[totalheight=6cm,width=7cm]{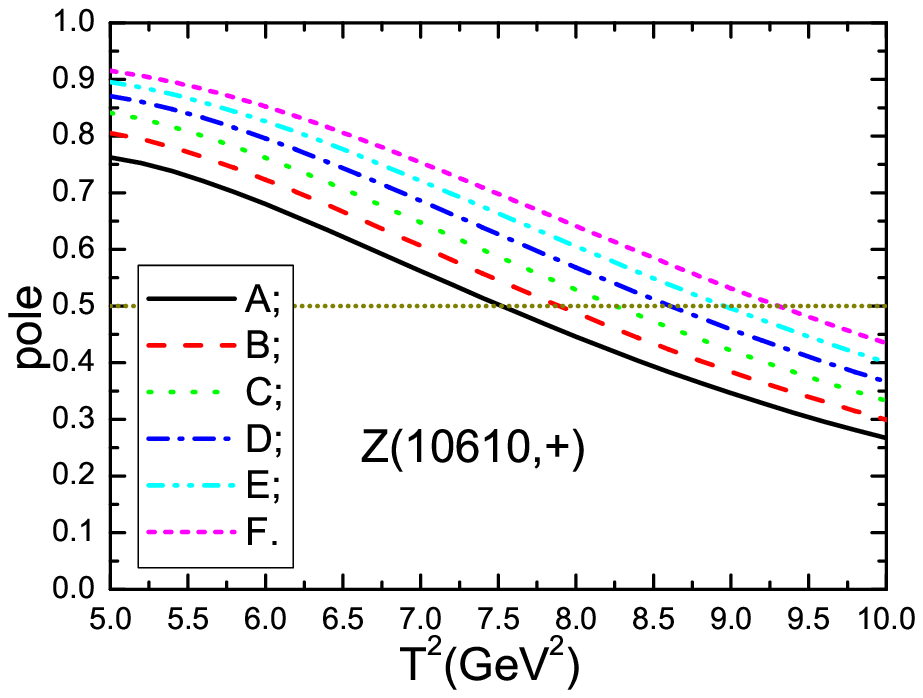}
\includegraphics[totalheight=6cm,width=7cm]{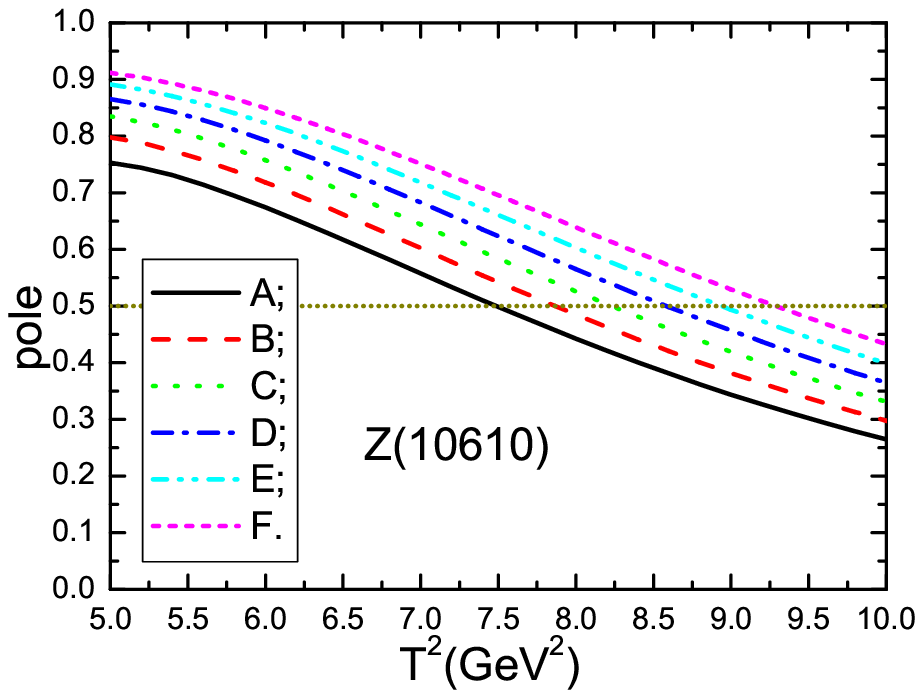}
\includegraphics[totalheight=6cm,width=7cm]{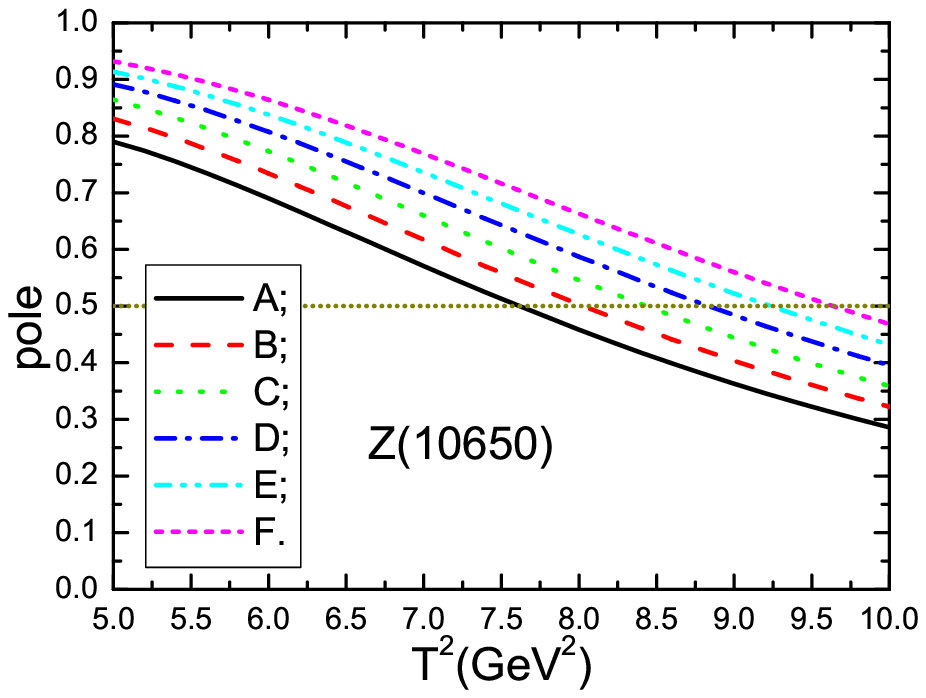}
  \caption{ The pole contributions  with variations of the  Borel parameters $T^2$ and threshold parameters $s_0$, where the $A$, $B$, $C$, $D$, $E$, $F$ denote the threshold parameters $s_0=120$,  $122$, $124$, $126$, $128$, $130\,\rm{GeV}^2$ respectively for the type I tetraquark states; $s_0=121$,  $123$, $125$, $127$, $129$, $131\,\rm{GeV}^2$ respectively for the type II tetraquark states; the $Z(10610,+)$ denotes the positive charge conjugation partner of the $Z_b(10610)$. }
\end{figure}

We take into account all uncertainties of the input parameters (including the vacuum condensates, the $b$-quark mass, the continuum threshold parameter, the energy scale and the Borel parameter) and
  obtain the values of the masses and pole residues of
 the   axial-vector hidden bottom tetraquark states, which are  shown explicitly in Figs.4-5 and Table 1.
 In this article,  we calculate the uncertainties $\delta$  with the
formula,
\begin{eqnarray}
\delta=\sqrt{\sum_i\left(\frac{\partial f}{\partial
x_i}\right)^2\mid_{x_i=\bar{x}_i} (x_i-\bar{x}_i)^2}\,  ,
\end{eqnarray}
 where the $f$ denotes   the  masses   and  pole residues  of the tetraquark states,  the $x_i$ denote the   input parameters $s_0$,  $T^2$, $\mu$, $m_b$, $\langle \bar{q}q \rangle$, $\langle \bar{q}g_s\sigma Gq \rangle$,  $\cdots$. As the partial  derivatives   $\frac{\partial f}{\partial x_i}$ are difficult to carry
out analytically, we take the  approximation $\left(\frac{\partial f}{\partial x_i}\right)^2 (x_i-\bar{x}_i)^2\approx
\left[f(\bar{x}_i\pm \Delta x_i)-f(\bar{x}_i)\right]^2$ in  numerical calculations with $x_i=\bar{x}_i\pm \Delta x_i$. From Table 1, we can see that the uncertainties of the masses
$M_Z$ are  about $1\%$,  while the uncertainties of the pole residues $\lambda_{Z}$ are about $15\%$.  We obtain the  squared masses   $M_Z^2$ through a
fraction, see Eq.(13), the uncertainties in the numerator and denominator which originate from a given input parameter (for example,
$\langle \bar{q}q \rangle$) cancel out with each other, and result in small net uncertainty.

\begin{table}
\begin{center}
\begin{tabular}{|c|c|c|c|c|c|c|c|}\hline\hline
 $J^{PC}$                &$T^2 (\rm{GeV}^2)$ &$s_0 (\rm{GeV}^2)$ &pole         &$M_{Z}(\rm{GeV})$        &$\lambda_{Z}$ \\ \hline
 $1^{++}$                &$7-8$              &$124\pm2$          &$(49-69)\%$  &$10.60^{+0.12}_{-0.09}$  &$1.40^{+0.23}_{-0.18}\times10^{-1}\rm{GeV}^5$  \\ \hline
 $1^{+-}$ ($Z_b(10610)$) &$7-8$              &$124\pm2$          &$(48-68)\%$  &$10.61^{+0.13}_{-0.09}$  &$1.42^{+0.24}_{-0.19}\times10^{-1}\rm{GeV}^5$  \\ \hline
 $1^{+-}$ ($Z_b(10650)$) &$7-8$              &$125\pm2$          &$(50-70)\%$  &$10.64^{+0.09}_{-0.08}$  &$1.72^{+0.24}_{-0.22}\times10^{-2}\rm{GeV}^4$ \\ \hline
\end{tabular}
\end{center}
\caption{ The Borel parameters, continuum threshold parameters, pole contributions, masses and pole residues of the  axial-vector  tetraquark states. }
\end{table}

The present  predictions  $M_{Z_b(10610)}=\left(10.61^{+0.11}_{-0.09}\right)\,\rm{GeV}$ and $M_{Z_b(10650)}=\left(10.64^{+0.08}_{-0.08}\right)\,\rm{GeV}$ are consistent with the experimental values  $ M_{Z_b(10610)}=\left(10607.2\pm2.0\right)\,\rm{ MeV}$  and $M_{Z_b(10650)}=\left(10652.2\pm1.5\right)\,\rm{MeV}$ \cite{Belle1110}.
The  predicted masses favor assigning  the $Z_b(10610)$ and $Z_b(10650)$  as the  $1^{+-}$   type I and type II tetraquark states, respectively. There is no candidate experimentally  for the $J^{PC}=1^{++}$ hidden bottom tetraquark states at the present time, the  prediction $M_Z=\left(10.60^{+0.11}_{-0.09}\right)\,\rm{GeV}$ can be confronted with the experimental data in the future at the LHCb and Belle-II.  The $C=+$ and $C=-$ type I axial-vector hidden bottom tetraquark states have degenerate  masses from the QCD sum rules.

 In the following, we perform Fierz re-arrangement  to the axial-vector  currents  both in the color and Dirac-spinor  spaces to   obtain the results,
\begin{eqnarray}
J_{1^{+-}}^{\mu}&=&\frac{\epsilon^{ijk}\epsilon^{imn}}{\sqrt{2}}\left\{u^jC\gamma_5 b^k \bar{d}^m\gamma^\mu C \bar{b}^n-u^jC\gamma^\mu b^k\bar{d}^m\gamma_5 C \bar{b}^n \right\} \, , \nonumber\\
 &=&\frac{1}{2\sqrt{2}}\left\{\,i\bar{b}i\gamma_5 b\,\bar{d}\gamma^\mu u-i\bar{b} \gamma^\mu b\,\bar{d}i\gamma_5 u+\bar{b} u\,\bar{d}\gamma^\mu\gamma_5 b-\bar{b} \gamma^\mu \gamma_5u\,\bar{d}b\right. \nonumber\\
&&\left. - i\bar{b}\gamma_\nu\gamma_5b\, \bar{d}\sigma^{\mu\nu}u+i\bar{b}\sigma^{\mu\nu}b\, \bar{d}\gamma_\nu\gamma_5u
- i \bar{b}\sigma^{\mu\nu}\gamma_5u\,\bar{d}\gamma_\nu b+i\bar{b}\gamma_\nu u\, \bar{d}\sigma^{\mu\nu}\gamma_5b   \,\right\} \, , \\
J_{1^{+-}}^{\mu\nu}&=&\frac{\epsilon^{ijk}\epsilon^{imn}}{\sqrt{2}}\left\{u^jC\gamma^\mu b^k \bar{d}^m\gamma^\nu C \bar{b}^n-u^jC\gamma^\nu b^k\bar{d}^m\gamma^\mu C \bar{b}^n \right\} \, , \nonumber\\
 &=&\frac{1}{2\sqrt{2}}\left\{\,i\bar{d}u\, \bar{b}\sigma^{\mu\nu}b +i\bar{d}\sigma^{\mu\nu}u \,\bar{b}b+i\bar{d}b\, \bar{b}\sigma^{\mu\nu}u +i\bar{d}\sigma^{\mu\nu}b \,\bar{b}u \right. \nonumber\\
 &&-\bar{b}\sigma^{\mu\nu}\gamma_5b\,\bar{d}i\gamma_5u-\bar{b}i\gamma_5 b\,\bar{d}\sigma^{\mu\nu}\gamma_5u -\bar{b}\sigma^{\mu\nu}\gamma_5u\,\bar{d}i\gamma_5b-\bar{d}i\gamma_5 b\,\bar{b}\sigma^{\mu\nu}\gamma_5u\nonumber\\
 &&+i\epsilon^{\mu\nu\alpha\beta}\bar{b}\gamma^\alpha\gamma_5b\, \bar{d}\gamma^\beta u-i\epsilon^{\mu\nu\alpha\beta}\bar{b}\gamma^\alpha b\, \bar{d}\gamma^\beta \gamma_5u\nonumber\\
 &&\left.+i\epsilon^{\mu\nu\alpha\beta}\bar{b}\gamma^\alpha\gamma_5u\, \bar{d}\gamma^\beta b-i\epsilon^{\mu\nu\alpha\beta}\bar{b}\gamma^\alpha u\, \bar{d}\gamma^\beta \gamma_5b \,\right\} \, , \\
J_{1^{++}}^{\mu}&=&\frac{\epsilon^{ijk}\epsilon^{imn}}{\sqrt{2}}\left\{u^jC\gamma_5 b^k \bar{d}^m\gamma^\mu C \bar{b}^n+u^jC\gamma^\mu b^k\bar{d}^m\gamma_5 C \bar{b}^n \right\} \, , \nonumber\\
 &=&\frac{1}{2\sqrt{2}}\left\{\,\bar{b}\gamma^\mu\gamma_5 b\,\bar{d} u-\bar{b} b\,\bar{d}\gamma^\mu\gamma_5 u+i\bar{b} \gamma^\mu u\,\bar{d}i\gamma_5 b-i\bar{b} i \gamma_5u\,\bar{d}\gamma^\mu b\right. \nonumber\\
&&\left. - i\bar{b}\gamma_\nu b\, \bar{d}\sigma^{\mu\nu}\gamma_5u+i\bar{b}\sigma^{\mu\nu}\gamma_5b\, \bar{d}\gamma_\nu u
- i \bar{b}\sigma^{\mu\nu}u\,\bar{d}\gamma_\nu\gamma_5 b+i\bar{b}\gamma_\nu\gamma_5 u\, \bar{d}\sigma^{\mu\nu}b   \,\right\} \, ,
\end{eqnarray}
where we add the subscripts $1^{+-}$ and $1^{++}$ to denote the $J^{PC}$ explicitly. Then we obtain the Okubo-Zweig-Iizuka super-allowed strong decays by taking into account the couplings to the meson-meson pairs,
\begin{eqnarray}
Z^{\pm}_b(10610)(1^{+-}) &\to& h_b({\rm 1P,2P})\pi^{\pm}\, , \, \Upsilon({\rm 1S,2S,3S})\pi^{\pm}\, , \, \eta_b({\rm 1S}) \rho^{\pm} \, , \, \eta_b({\rm 1S,2S})(\pi\pi)_{\rm P}^{\pm} \, ,\nonumber\\
Z^{\pm}_b(10650)(1^{+-}) &\to&  \Upsilon({\rm 1S,2S,3S})\pi^{\pm}\, , \, \eta_b({\rm 1S}) \rho^{\pm}\, , \, \eta_b({\rm 1S,2S})(\pi\pi)_{\rm P}^{\pm}\, , \, \chi_{b1}({\rm 1P,2P})(\pi\pi)_{\rm P}^{\pm} \, , \, (B\bar{B}^*)^\pm \, ,\nonumber\\
Z^{\pm}_b(10600)(1^{++}) &\to&  \chi_{b0}({\rm 1P,2P})\pi^{\pm}\, , \, \Upsilon({\rm 1S}) \rho^{\pm}\, , \, \Upsilon({\rm 1S,2S})(\pi\pi)_{\rm P}^{\pm}\, ,
\end{eqnarray}
where we use the $(\pi\pi)_{\rm P}$ to denote  the P-wave $\pi\pi$ systems have the same quantum numbers of the $\rho$,  and take the decays to the $(\pi\pi)_{\rm P}^{\pm}$   final states as Okubo-Zweig-Iizuka super-allowed according to the decays $\rho \to \pi\pi$.
 In this article, we  denote the hidden bottom tetraquark states with the mass $10600\,\rm{MeV}$ as the $Z_b(10600)$, see Table 1.
We can search for the $Z^{\pm}_b(10650)(1^{+-})$ in the typical decays,
\begin{eqnarray}
Z^{\pm}_b(10650)(1^{+-}) &\to&   \chi_{b1}({\rm 1P,2P})(\pi\pi)_{\rm P}^{\pm} \, , \, (B\bar{B}^*)^\pm \, ,
\end{eqnarray}
which originate from the typical sub-structures of the $Z^{\pm}_b(10650)(1^{+-})$.

In the nonrelativistic and heavy quark limit, the components $\bar{b}\sigma^{\mu\nu}\gamma_5u\,\bar{d}\gamma_\nu b$ and $\epsilon^{\mu\nu\alpha\beta}\bar{b}\gamma^\alpha\gamma_5u\, \bar{d}\gamma^\beta b$ of the interpolating  currents $J_{1^{+-}}^{\mu}$ and $J_{1^{+-}}^{\mu\nu}$  respectively  are reduced to the following forms,
\begin{eqnarray}
\bar{b}\sigma^{0j}\gamma_5 u \, \bar{d}\gamma_j b&\propto&  \xi^{\dagger}_b\sigma^j\zeta_u\, \chi^{\dagger}_d\vec{\sigma}\cdot \vec{k}_{d}\sigma^j\xi_b\,\,\,\propto\,\,\, \xi^{\dagger}_b\frac{\sigma^j}{2}\zeta_u\, \chi^{\dagger}_d\frac{\sigma^j}{2}\xi_b=\vec{S}_{B^*} \cdot \vec{S}_{\bar{B}^*}  \, , \nonumber\\
\bar{b}\sigma^{ij}\gamma_5 u \, \bar{d}\gamma_j b&\propto& \epsilon^{ijk} \xi^{\dagger}_b\sigma^k\vec{\sigma}\cdot \vec{k}_{u}\zeta_u\, \chi^{\dagger}_d\vec{\sigma}\cdot \vec{k}_{d}\sigma^j\xi_b\,\,\, \propto\,\,\, \epsilon^{ijk} \xi^{\dagger}_b\frac{\sigma^k}{2}\zeta_u\, \chi^{\dagger}_d\frac{ \sigma^j}{2}\xi_b=\vec{S}_{\bar{B}^*}\times \vec{S}_{B^*} \, , \nonumber\\
\epsilon^{ijk}\bar{b}\gamma^{j}\gamma_5 u \, \bar{d}\gamma^k  b&\propto& \epsilon^{ijk} \xi^{\dagger}_b\sigma^j\zeta_u\, \chi^{\dagger}_d\vec{\sigma}\cdot \vec{k}_{d}\sigma^k\xi_b\,\,\, \propto\,\,\, \epsilon^{ijk} \xi^{\dagger}_b\frac{\sigma^j}{2}\zeta_u\, \chi^{\dagger}_d\frac{ \sigma^k}{2}\xi_b=\vec{S}_{B_1}\times \vec{S}_{\bar{B}^*} \, ,\nonumber \\
\epsilon^{ijk}\bar{b}\gamma^{0}\gamma_5 u \, \bar{d}\gamma^k  b&\propto& \epsilon^{ijk} \xi^{\dagger}_b\vec{\sigma}\cdot \vec{k}_{u}\zeta_u\, \chi^{\dagger}_d\vec{\sigma}\cdot \vec{k}_{d}\sigma^k\xi_b\,\,\, \propto\,\,\, \epsilon^{ijk} \xi^{\dagger}_b\zeta_u\, \chi^{\dagger}_d\frac{ \sigma^k}{2}\xi_b=\epsilon^{ijk} S^k_{\bar{B}^*} \, ,
\end{eqnarray}
where the $\xi$, $\zeta$, $\chi$ are the two-component spinors of the  quark fields, the $\vec{k}$  are the three-vectors of the  quark fields,   the $\sigma^i$ are the pauli matrixes, and the $\vec{S}$ are the spin operators.
The thresholds are  $B^*\bar{B}^*=10650\,\rm{MeV}$,  $B\bar{B}^*=10605\,\rm{MeV}$, $B_0\bar{B}^*\approx B_1\bar{B}^*=11049\,\rm{MeV}$ \cite{PDG}.
It is obvious that the currents $\bar{b}\sigma^{\mu\nu}\gamma_5u\,\bar{d}\gamma_\nu b$ and $\bar{b}\gamma_\nu u\, \bar{d}\sigma^{\mu\nu}\gamma_5b$ ($\epsilon^{\mu\nu\alpha\beta}\bar{b}\gamma^\alpha\gamma_5u\, \bar{d}\gamma^\beta b$ and $\epsilon^{\mu\nu\alpha\beta}\bar{b}\gamma^\alpha u\, \bar{d}\gamma^\beta \gamma_5b$)
couple   to the $J^P=0^+$ and $1^+$ $(B^*\bar{B}^*)^+$ ($J^P=1^+$  $(B_1^*\bar{B}^*)^+$ and $(B_0^*\bar{B}^*)^+$) states. The strong decays
\begin{eqnarray}
Z^{\pm}_b(10610)(1^{+-}) &\to& (B^*\bar{B}^*)^{\pm} \, ,\nonumber\\
Z^{\pm}_b(10650)(1^{+-}) &\to& (B_1\bar{B}^*)^{\pm} \, ,
\end{eqnarray}
are Okubo-Zweig-Iizuka super-allowed but kinematically forbidden.
The $Z^{\pm}_b(10610)$ and $Z^{\pm}_b(10650)$ have the same quantum numbers and analogous  strong decays but different masses and quark configurations.

Now we list out the possible strong decays of the $Z^{\pm}_b(10610)$, $Z^{\pm}_b(10650)$ and $Z_b^{\pm}(10600)$,
\begin{eqnarray}
Z^{\pm}_b(10610)(1^{+-}) &\to& h_b({\rm 1P,2P})\pi^{\pm}\, , \, \Upsilon({\rm 1S,2S,3S})\pi^{\pm}\, , \, \eta_b({\rm 1S}) \rho^{\pm} \, , \, \eta_b({\rm 1S,2S})(\pi\pi)_{\rm P}^{\pm}\, , \,\nonumber\\
&& \chi_{b1}({\rm 1P,2P})(\pi\pi)_{\rm P}^{\pm}\, ,\nonumber\\
Z^{\pm}_b(10650)(1^{+-}) &\to&  h_b({\rm 1P,2P})\pi^{\pm}\, , \, \Upsilon({\rm 1S,2S,3S})\pi^{\pm}\, , \, \eta_b({\rm 1S}) \rho^{\pm} \, , \,\eta_b({\rm 1S,2S})(\pi\pi)_{\rm P}^{\pm}\, , \,\nonumber\\
&&  \chi_{b1}({\rm 1P,2P})(\pi\pi)_{\rm P}^{\pm} \, , \, (B\bar{B}^*)^\pm\, , \, (B^*{\bar B}^*)^\pm \, ,\nonumber\\
Z^{\pm}_b(10600)(1^{++}) &\to&  \chi_{b0}({\rm 1P,2P})\pi^{\pm}\, , \, \chi_{b1}({\rm 1P,2P})\pi^{\pm}\, , \,\Upsilon({\rm 1S}) \rho^{\pm}\, , \, \Upsilon({\rm 1S,2S})(\pi\pi)_{\rm P}^{\pm}\, .
\end{eqnarray}

 The following strong decays take place through the re-scattering mechanism,
\begin{eqnarray}
Z^{\pm}_b(10610)(1^{+-}) &\to&\chi_{b1}({\rm 1P,2P})(\pi\pi)_{\rm P}^{\pm}\, ,\nonumber\\
Z^{\pm}_b(10650)(1^{+-}) &\to&  h_b({\rm 1P,2P})\pi^{\pm}\, , \, (B^*{\bar B}^*)^\pm \, ,\nonumber\\
Z^{\pm}_b(10600)(1^{++}) &\to&   \chi_{b1}({\rm 1P,2P})\pi^{\pm}\, ,
\end{eqnarray}
and cannot be the dominant decay modes.

We can also search for the neutral partner $Z^{0}_b(10610/10650)(1^{+-})$
in the following strong and electromagnetic decays,
\begin{eqnarray}
Z^{0}_b(10610/10650)(1^{+-}) &\to& h_c({\rm 1P,2P})\pi^{0}  \, , \,\Upsilon({\rm 1S,2S,3S})\pi^{0}\, , \, \eta_b ({\rm 1S})\rho^{0}\, , \, \eta_b ({\rm 1S}) \omega\, , \, \eta_b ({\rm 1S,2S})(\pi\pi)_{\rm P}^{0}  \, , \,  \nonumber\\
&& \chi_{bj}({\rm 1P,2P})(\pi\pi)_{\rm P}^{0}\, , \,\eta_b ({\rm 1S,2S})(\pi\pi\pi)_{\rm P}^{0} \, , \, \chi_{bj} ({\rm 1P})(\pi\pi\pi)_{\rm P}^{0}\, , \,\eta_b ({\rm 1S,2S}) \gamma \, , \,\nonumber\\
&& \chi_{bj} ({\rm 1P,2P}) \gamma\, , \, (B\bar{B}^*)^0\, ,
\end{eqnarray}
where the $(\pi\pi\pi)_{\rm P}$ denotes the P-wave $\pi\pi\pi$ systems with the same quantum numbers of the $\omega$.

 The diquark-antidiquark type current with special quantum numbers couples    to a special tetraquark state, while the current can be re-arranged both in the color and Dirac-spinor  spaces, and changed  to a current as a special superposition of   color  singlet-singlet type currents.   The color  singlet-singlet type currents couple to the meson-meson pairs. The
diquark-antidiquark type tetraquark state can be taken as a special superposition of a series of  meson-meson pairs, and embodies  the net effects. The decays to its components (meson-meson pairs) are Okubo-Zweig-Iizuka super-allowed, but the re-arrangements in the color-space are non-trivial \cite{Nielsen3900}.

\begin{figure}
\centering
\includegraphics[totalheight=6cm,width=7cm]{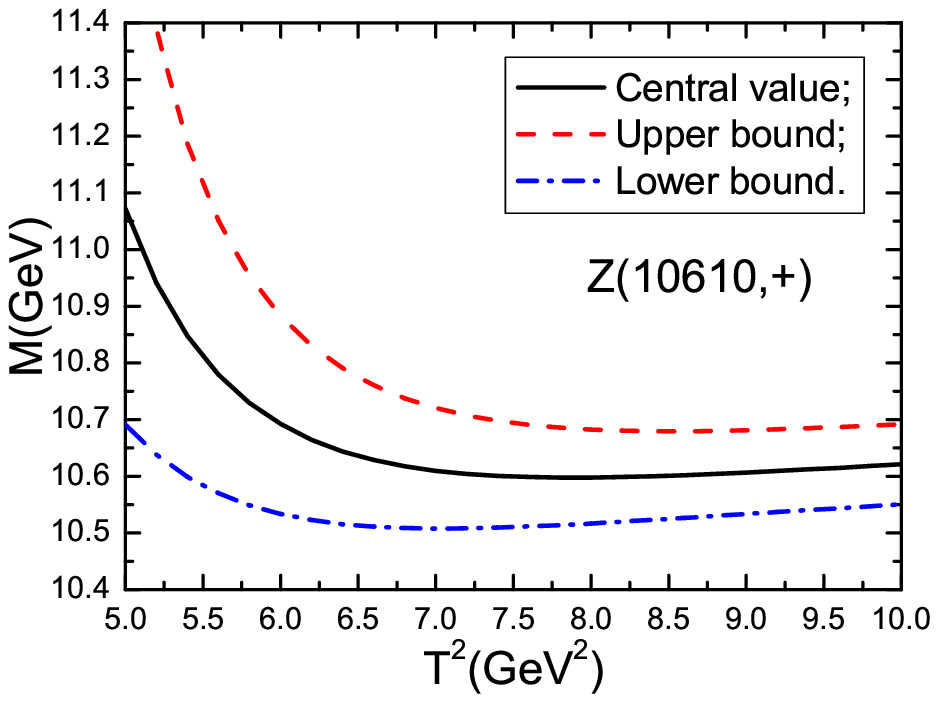}
\includegraphics[totalheight=6cm,width=7cm]{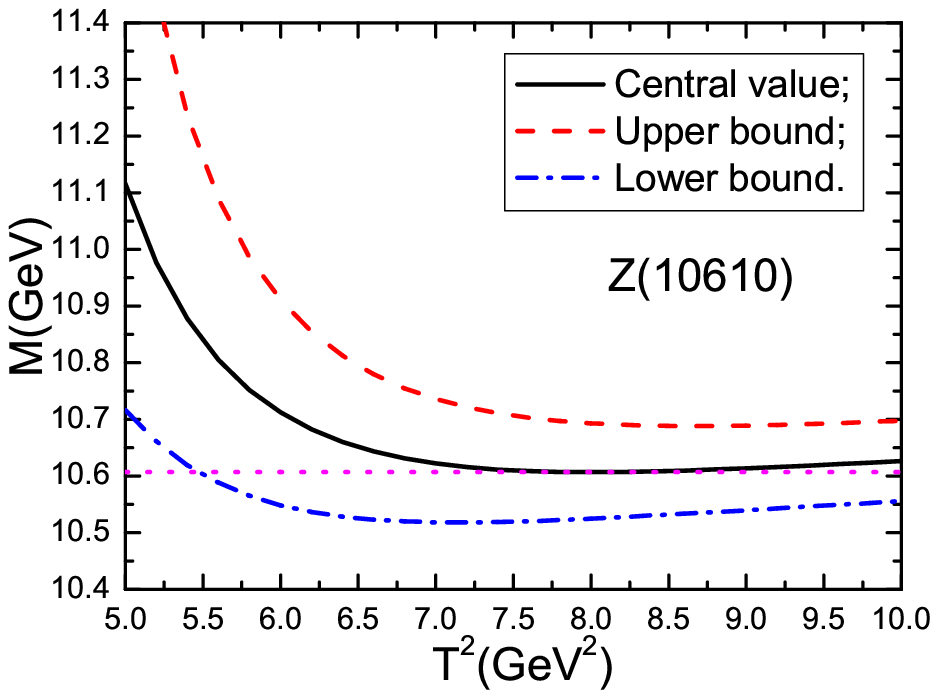}
\includegraphics[totalheight=6cm,width=7cm]{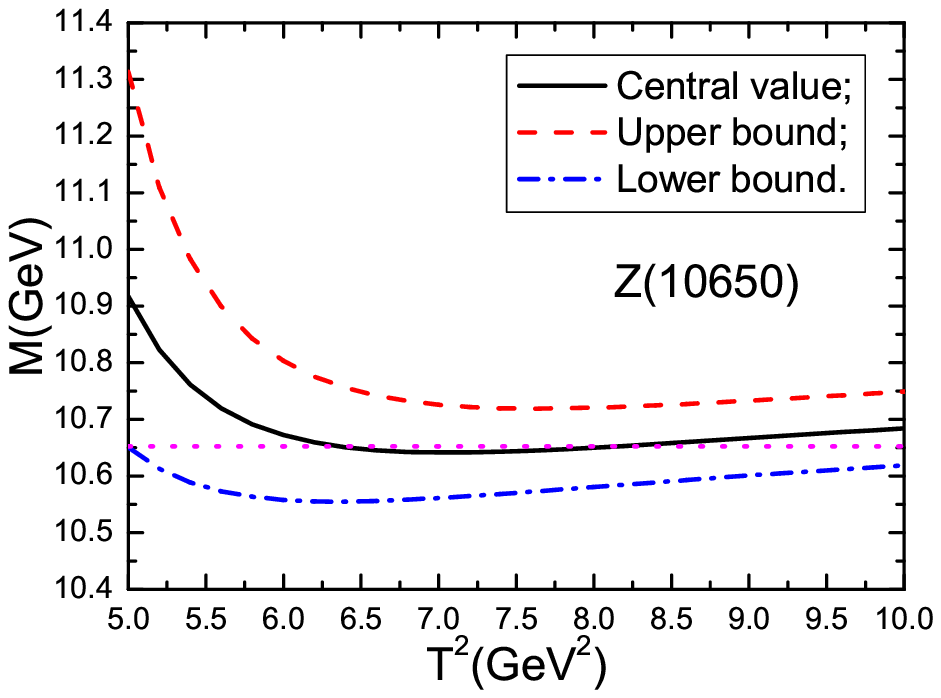}
  \caption{ The masses  with variations of the  Borel parameters $T^2$, where the horizontal lines denote  the experimental values, the $Z(10610,+)$ denotes the positive charge conjugation partner of the $Z_b(10610)$.}
\end{figure}

\begin{figure}
\centering
\includegraphics[totalheight=6cm,width=7cm]{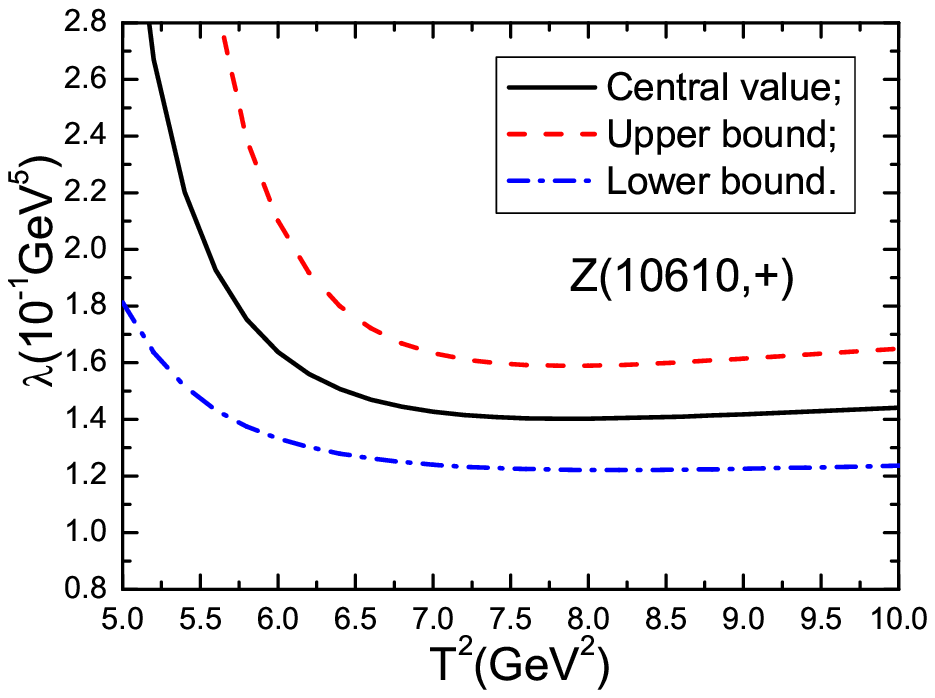}
\includegraphics[totalheight=6cm,width=7cm]{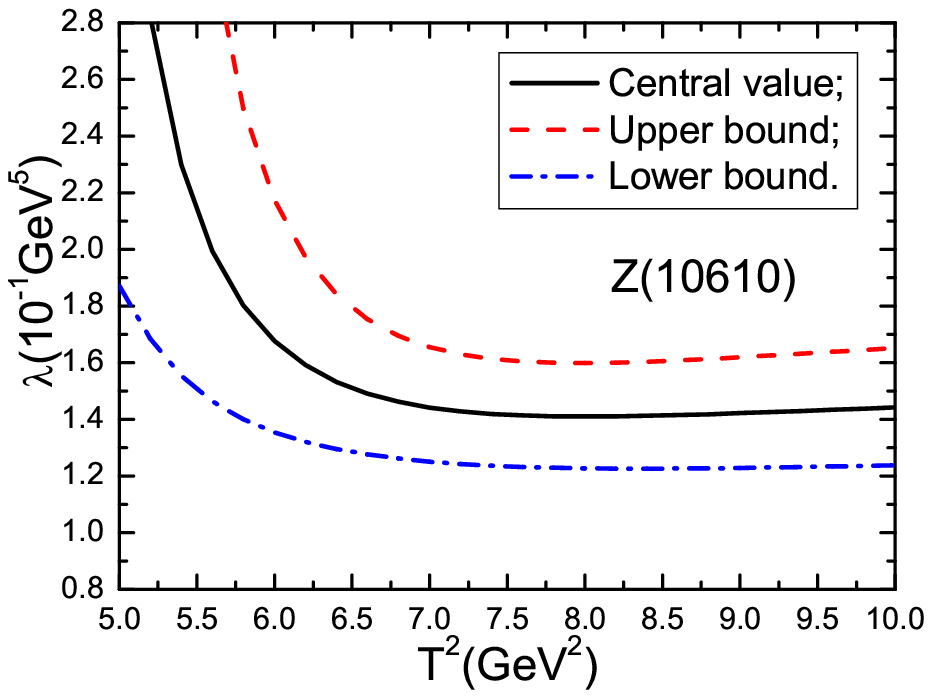}
\includegraphics[totalheight=6cm,width=7cm]{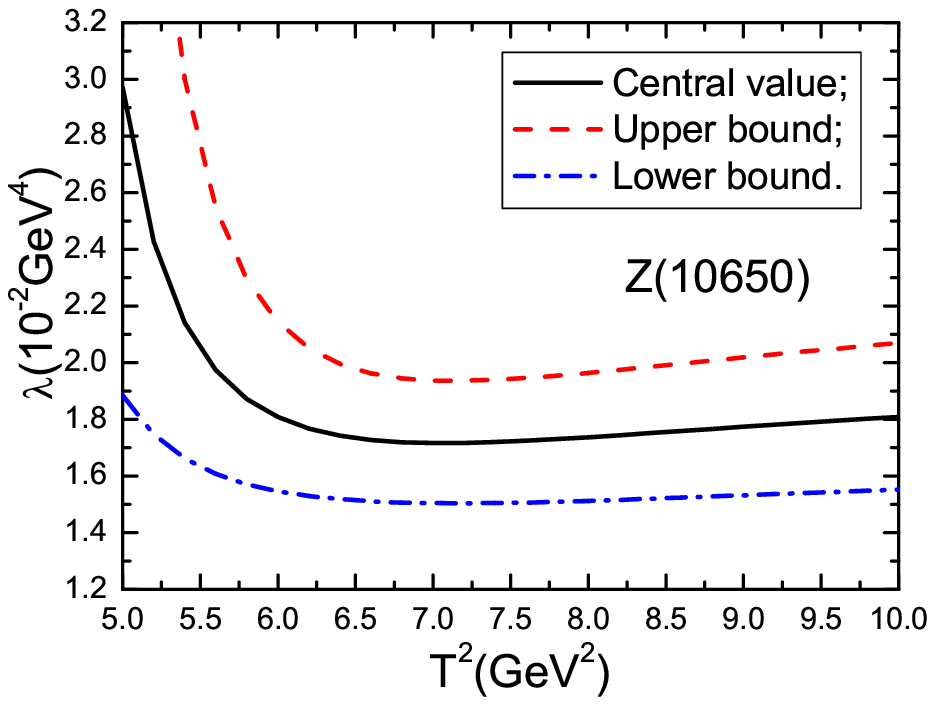}
  \caption{ The pole residues   with variations of the  Borel parameters $T^2$, where the $Z(10610,+)$ denotes the positive charge conjugation  partner of the $Z_b(10610)$. }
\end{figure}

\section{Strong decays $Z_b^\pm(10610)\to\Upsilon\pi^{\pm}, \, \, \eta_b\rho^{\pm}$}

The  pole residues $\lambda_{Z_b}$ can be taken as   basic input parameters to study relevant processes of the  axial-vector  tetraquark states $Z^{\pm}_b(10610)$, $Z^{\pm}_b(10650)$ and $Z_b^{\pm}(10600)$ with the three-point QCD sum rules. For example, we can study the strong decays $Z_b^\pm(10610)\to\Upsilon\pi^{\pm}$ and $\eta_b\rho^{\pm}$ with the following three-point correlation functions
$\Pi_{\mu,\nu}^{1}(p,q)$ and $\Pi_{\mu,\nu}^{2}(p,q)$, respectively,
\begin{eqnarray}
\Pi_{\mu,\nu}^{1}(p,q)&=&i^2\int d^4xd^4y e^{ipx}e^{iqy}\langle 0|T\left\{J_\mu^{\Upsilon}(x)J_5^{\pi}(y)J_{\nu,1^{+-}}(0)\right\}|0\rangle\, , \nonumber \\
\Pi_{\mu,\nu}^{2}(p,q)&=&i^2\int d^4xd^4y e^{ipx}e^{iqy}\langle 0|T\left\{J_5^{\eta_b}(x)J_\mu^{\rho}(y)J_{\nu,1^{+-}}(0)\right\}|0\rangle \, ,
\end{eqnarray}
where the currents
\begin{eqnarray}
J_\mu^{\Upsilon}(x)&=&\bar{b}(x)\gamma_\mu b(x) \, ,\nonumber \\
J_\mu^{\rho}(y)&=&\bar{u}(y)\gamma_\mu d(y) \, ,\nonumber \\
J_5^{\eta_b}(x)&=&\bar{b}(x)i\gamma_5 b(x) \, ,\nonumber \\
J_5^{\pi}(y)&=&\bar{u}(y)i\gamma_5 d(y) \, ,
\end{eqnarray}
interpolate the mesons $\Upsilon$, $\rho$, $\eta_b$, $\pi$, respectively.

We insert  a complete set of intermediate hadronic states with
the same quantum numbers as the current operators into the three-point
correlation functions  and  isolate the ground state
contributions to obtain the following results,
\begin{eqnarray}
\Pi_{\mu,\nu}^{1}(p,q)&=& \frac{f_{\pi}M_{\pi}^2f_{\Upsilon}M_{\Upsilon}\lambda_{Z_b}G_{Z_b\Upsilon \pi}}{m_u+m_d} \frac{-i}{(M_{Z_b}^2-p^{\prime2})(M_{\Upsilon}^2-p^2)(M_{\pi}^2-q^2)} \left(-g_{\mu\alpha}+\frac{p_{\mu}p_{\alpha}}{p^2} \right) \nonumber\\
&&\left(-g_{\nu}^{\alpha}+\frac{p^{\prime}_{\nu}p^{\prime\alpha}}{p^{\prime2}} \right)+\cdots \, , \nonumber\\
\Pi_{\mu,\nu}^{2}(p,q)&=& \frac{f_{\eta_b}M_{\eta_b}^2f_{\rho}M_{\rho}\lambda_{Z_b}G_{Z_b\eta_b \rho}}{2m_b} \frac{-i}{(M_{Z_b}^2-p^{\prime2})(M_{\eta_b}^2-p^2)(M_{\rho}^2-q^2)} \left(-g_{\mu\alpha}+\frac{q_{\mu}q_{\alpha}}{q^2} \right) \nonumber\\
&&\left(-g_{\nu}^{\alpha}+\frac{p^{\prime}_{\nu}p^{\prime\alpha}}{p^{\prime2}} \right)+\cdots \, ,
\end{eqnarray}
where $p^\prime=p+q$, the $f_\Upsilon$, $f_{\eta_b}$, $f_{\rho}$ and $f_{\pi}$ are the decay constants of the mesons  $\Upsilon$, $\eta_b$, $\rho$ and $\pi$, respectively, the $G_{Z_b\Upsilon\pi}$ and $G_{Z_b\eta_b\rho}$ are the hadronic coupling constants. In the following, we write down the definitions,
\begin{eqnarray}
\langle0|J_{\mu}^\Upsilon(0)|\Upsilon(p)\rangle&=&f_{\Upsilon}M_{\Upsilon}\xi_\mu \,\, , \nonumber \\
\langle0|J_{\mu}^\rho(0)|\rho(q)\rangle&=&f_{\rho}M_{\rho}\varepsilon_\mu \,\, , \nonumber \\
\langle0|J_{5}^{\eta_b}(0)|\eta_b(p)\rangle&=&\frac{f_{\eta_b}M_{\eta_b}^2}{2m_b} \,\, , \nonumber \\
\langle0|J_{5}^{\pi}(0)|\pi(q)\rangle&=&\frac{f_{\pi}M_{\pi}^2}{m_u+m_d} \,\, ,\nonumber \\
\langle\Upsilon(p)\pi(q)|Z_b(p^{\prime})\rangle&=&\xi^*(p)\cdot\zeta(p^{\prime}) G_{Z_b\Upsilon\pi}(q^2) \, , \nonumber\\
\langle\eta_b(p)\rho(q)|Z_b(p^{\prime})\rangle&=&\varepsilon^*(q)\cdot\zeta(p^{\prime}) G_{Z_b\eta_b\rho}(q^2) \, ,
\end{eqnarray}
the $\xi$, $\zeta$ and $\varepsilon$ are polarization vectors of the $\Upsilon$, $Z_b$ and $\rho$, respectively.
Now we choose the tensors $q_{\mu}p_\nu$ and $p_{\mu}q_\nu$ to study the  coupling constants $G_{Z_b\Upsilon\pi}$ and $G_{Z_b\eta_b\rho}$, respectively.
We carry out the operator product expansion and take into account the color connected Feynman diagrams \cite{Nielsen3900},
\begin{eqnarray}
\Pi^{1}_{\mu\nu}(p,q)&=&\frac{i m_b\langle \bar{q}g_s\sigma Gq\rangle q_{\mu}p_\nu}{48\sqrt{2}\pi^2q^2}\int_0^1 dx \frac{1}{x(1-x)p^2-m_b^2} \nonumber\\
&&+\frac{ig_s^2\langle\bar{q}q\rangle^2q_{\mu}p_\nu}{81\sqrt{2}\pi^2q^2}\int_0^1 dx \left\{ \frac{3}{2\left[x(1-x)p^2-m_b^2\right]} +\frac{3x(1-x)m_b^2}{2\left[x(1-x)p^2-m_b^2\right]^2}\right.\nonumber\\
&&\left.- \frac{4x(1-x)}{x(1-x)p^2-m_b^2}-\frac{\left[x^2+(1-x)^2\right]m_b^2}{\left[x(1-x)p^2-m_b^2\right]^2} \right\} \, ,
\end{eqnarray}
\begin{eqnarray}
\Pi^{2}_{\mu\nu}(p,q)&=&-\frac{i m_b\langle \bar{q}g_s\sigma Gq\rangle p_{\mu}q_\nu}{48\sqrt{2}\pi^2q^2}\int_0^1 dx \frac{1}{x(1-x)p^2-m_b^2} \nonumber\\
&&-\frac{ig_s^2\langle\bar{q}q\rangle^2p_{\mu}q_\nu}{81\sqrt{2}\pi^2q^2}\int_0^1 dx \left\{ \frac{3}{2\left[x(1-x)p^2-m_b^2\right]} +\frac{3x(1-x)m_b^2}{2\left[x(1-x)p^2-m_b^2\right]^2}\right.\nonumber\\
&&\left.- \frac{4x(1-x)}{x(1-x)p^2-m_b^2}-\frac{\left[x^2+(1-x)^2\right]m_b^2}{\left[x(1-x)p^2-m_b^2\right]^2} \right\} \, .
\end{eqnarray}

Then we take  the Borel transform with respect to the variable   $P^2=-p^2=-p^{\prime2} $ and obtain the following QCD sum rules,
\begin{eqnarray}
&&\frac{f_{\pi}M_{\pi}^2f_{\Upsilon}M_{\Upsilon}\lambda_{Z_b}G_{Z_b\Upsilon \pi}}{(m_u+m_d)M_{Z_b}^2(M_{Z_b}^2-M_{\Upsilon}^2)} \left\{ \exp\left(-\frac{M_{\Upsilon}^2}{T^2} \right)-\exp\left(-\frac{M_{Z_b}^2}{T^2} \right)\right\}+C \exp\left(-\frac{s_0}{T^2} \right) \nonumber\\
&&=\frac{m_b\langle\bar{q}g_s\sigma Gq\rangle}{48\sqrt{2}\pi^2}\frac{Q^2+M_{\pi}^2}{Q^2}\int_0^1 dx \frac{1}{x(1-x)}\exp\left( -\frac{m_b^2}{x(1-x)T^2}\right)\nonumber\\
&&+\frac{g_s^2\langle\bar{q}q\rangle^2}{81\sqrt{2}\pi^2}\frac{Q^2+M_{\pi}^2}{Q^2}\int_0^1 dx \left\{\frac{3}{2x(1-x)}\left( 1-\frac{m_b^2}{T^2}\right)-4\left[1-\left(\frac{1}{x^2}+\frac{1}{(1-x)^2} \right)\frac{m_b^2}{4T^2}\right]\right\}\nonumber\\
&&\exp\left( -\frac{m_b^2}{x(1-x)T^2}\right)\, ,
\end{eqnarray}
\begin{eqnarray}
&&\frac{f_{\eta_b}M_{\eta_b}^2f_{\rho}M_{\rho}\lambda_{Z_b}G_{Z_b\eta_b \rho}}{2m_b M_{Z_b}^2(M_{Z_b}^2-M_{\eta_b}^2)} \left\{ \exp\left(-\frac{M_{\eta_b}^2}{T^2} \right)-\exp\left(-\frac{M_{Z_b}^2}{T^2} \right)\right\}+C \exp\left(-\frac{s_0}{T^2} \right) \nonumber\\
&&=-\frac{m_b\langle\bar{q}g_s\sigma Gq\rangle}{48\sqrt{2}\pi^2}\frac{Q^2+M_{\rho}^2}{Q^2}\int_0^1 dx \frac{1}{x(1-x)}\exp\left( -\frac{m_b^2}{x(1-x)T^2}\right)\nonumber\\
&&-\frac{g_s^2\langle\bar{q}q\rangle^2}{81\sqrt{2}\pi^2}\frac{Q^2+M_{\rho}^2}{Q^2}\int_0^1 dx \left\{\frac{3}{2x(1-x)}\left( 1-\frac{m_b^2}{T^2}\right)-4\left[1-\left(\frac{1}{x^2}+\frac{1}{(1-x)^2} \right)\frac{m_b^2}{4T^2}\right]\right\}\nonumber\\
&&\exp\left( -\frac{m_b^2}{x(1-x)T^2}\right)\, ,
\end{eqnarray}
where the $s_0$ is the continuum threshold parameter for the $Z_b(10610)$, and the $C$ are unknown  parameters introduced to take into account
single-pole contributions associated with pole-continuum
transitions. In the three-point QCD sum rules, the single-pole contributions  are not suppressed if a single
Borel transform is taken.

The input parameters are taken as $M_{\pi}=0.140\,\rm{GeV}$, $f_{\pi}=0.130\,\rm{GeV}$,
$M_{\Upsilon}=9.4603\,\rm{GeV}$, $M_{\eta_b}=9.398\,\rm{GeV}$, $M_{\rho}=0.775\,\rm{GeV}$, $f_{\rho}=0.215\,\rm{GeV}$,
$f_{\Upsilon}=f_{\eta_b}=0.700 \,\rm{GeV}$ \cite{PDG,RCVerma}, and $m_u({\mu=\rm 1GeV})=m_d({\mu=\rm 1GeV})=0.006\,\rm{GeV}$ from the Gell-Mann-Oakes-Renner relation.
The unknown parameters are chosen as $C=0.0014\,\rm{GeV}^6 $ and $-0.0010\,\rm{GeV}^6 $ in the QCD sum rules for the coupling constants $G_{Z_b\Upsilon\pi}$ and $G_{Z_b\eta_b\rho}$ respectively to obtain  platforms in the Borel windows $T^2=(7-8)\,\rm{GeV}^2$. The central values of the $G_{Z_b\Upsilon\pi}$ and $G_{Z_b\eta_b\rho}$ can be fitted to the following forms,
\begin{eqnarray}
|G_{Z_b\Upsilon\pi}(Q^2)|&=&3.53\,\rm{GeV}\, , \nonumber \\
G_{Z_b\eta_b\rho}(Q^2)&=&\frac{1421.9\,{\rm GeV^3}}{257.4\,{\rm GeV^2}+Q^2} \, ,
\end{eqnarray}
with $Q^2=-q^2$.
We extend the coupling constants to the physical regions and take into account the uncertainties,
\begin{eqnarray}
|G_{Z_b\Upsilon\pi}\left(Q^2=-M_\pi^2\right)|&=&3.53^{+1.21}_{-0.91}\,\rm{GeV}\, , \nonumber\\
G_{Z_b\eta_b\rho}\left(Q^2=-M_\rho^2\right)&=&5.54^{+1.82}_{-1.42}\,\rm{GeV}\, .
\end{eqnarray}

The resulting  decay widths are
\begin{eqnarray}
\Gamma(Z_b^+(10610)\to\Upsilon\pi^+)&=& \frac{p\left(M_{Z_b},M_{\Upsilon},M_{\pi}\right)}{24\pi M_{Z_b}^2}G_{Z_b\Upsilon\pi}^2\left(3+\frac{p\left(M_{Z_b},M_{\Upsilon},M_{\pi}\right)^2}{M_{\Upsilon}^2} \right)\nonumber\\
&=&4.77^{+3.27}_{-2.46}\,\rm{MeV}   \, ,\nonumber\\
\Gamma(Z_b^+(10610)\to\eta_b\rho^+)&=& \frac{p\left(M_{Z_b},M_{\eta_b},M_{\rho}\right)}{24\pi M_{Z_b}^2}G_{Z_b\eta_b\rho}^2\left(3+\frac{p\left(M_{Z_b},M_{\eta_b},M_{\rho}\right)^2}{M_{\rho}^2} \right)\nonumber\\
&=&13.52^{+8.89}_{-6.93} \,\rm{MeV}  \, ,
\end{eqnarray}
where $p(a,b,c)=\frac{\sqrt{[a^2-(b+c)^2][a^2-(b-c)^2]}}{2a}$.	Those  widths are consistent with the experimental data  $\Gamma_{Z_b(10610)}=(18.4\pm2.4) \,\rm{MeV}$ from the Belle collaboration \cite{Belle1110}, the present calculations support  assigning the $Z_b(10610)$ as the $1^{+-}$ diquark-antidiquark type tetraquark state. We can search for the $Z_b^\pm(10610)$ in the final states $\eta_b\rho^\pm$.
The strong decays $Z^{\pm}_b(10610)(1^{+-})\to h_b({\rm 1P,2P})\pi^{\pm}$ take place through relative P-wave,  the decay widths
	$\Gamma(Z^{\pm}_b(10610)(1^{+-})\to h_b({\rm 1P,2P})\pi^{\pm})\propto p\left(M_{Z_b},M_{h_b},M_{\pi}\right)^3$, and the decays are kinematically suppressed in the phase-space. Detailed studies based on the QCD sum rules are postponed to our next work.

\section{Conclusion}
In this article, we study the axial-vector mesons $Z_b(10610)$ and $Z_b(10650)$ with the $C\gamma_\mu-C\gamma_5$ type and $C\gamma_\mu-C\gamma_\nu$ type interpolating currents respectively by carrying out the operator product expansion to  the vacuum condensates up to dimension-10.  In calculations, we study the energy scale dependence of the QCD spectral densities in details for the first time, and suggest a formula $\mu=\sqrt{M^2_{X/Y/Z}-(2{\mathbb{M}}_b)^2}$ with
 the effective mass ${\mathbb{M}}_b=5.13\,\rm{GeV}$ to determine  the energy scales, which works very well. The numerical results support assigning the $Z_b(10610)$ and $Z_b(10650)$ as the $C\gamma_\mu-C\gamma_5$ type and $C\gamma_\mu-C\gamma_\nu$ type hidden bottom tetraquark states,  respectively. The $Z_b(10610)$, $Z_b(10650)$, $Z_c(3900)$ and $Z_c(4020)$ are observed in the analogous decays to the final states $\pi^{\pm}\Upsilon({\rm 1,2,3S})$, $\pi^{\pm} h_b({\rm 1,2P})$, $\pi^\pm J/\psi$, $\pi^\pm h_c$, and should have analogous structures. Furthermore, we obtain the mass of the   $C\gamma_\mu-C\gamma_5$ type $J^{PC}=1^{++}$ hidden bottom tetraquark state, which can be confronted with the experimental data in the future at the LHCb and Belle-II. The  pole residues $\lambda_{Z_b}$ can be taken as   basic input parameters to study relevant processes of the  axial-vector  tetraquark states $Z^{\pm}_b(10610)$, $Z^{\pm}_b(10650)$ and $Z_b^{\pm}(10600)$ with the three-point QCD sum rules. We   study the strong decays $Z_b^\pm(10610)\to\Upsilon\pi^{\pm}\, ,\,\eta_b\rho^{\pm}$ with the three-point QCD sum rules, the decay widths also support assigning the $Z_b(10610)$ as the $C\gamma_\mu-C\gamma_5$ type hidden bottom tetraquark state.

\section*{Acknowledgements}
This  work is supported by National Natural Science Foundation,
Grant Numbers 11375063, 11235005,    the Fundamental Research Funds for the
Central Universities, and Natural Science Foundation of Hebei province, Grant Number A2014502017.

\section*{Appendix}
The spectral densities at the level of the quark-gluon degrees of
freedom,

\begin{eqnarray}
\rho^{\rm I}_{0}(s)&=&\frac{1}{3072\pi^6}\int_{y_i}^{y_f}dy \int_{z_i}^{1-y}dz \, yz(1-y-z)^3\left(s-\overline{m}_b^2\right)^2\left(35s^2-26s\overline{m}_b^2+3\overline{m}_b^2 \right)\, ,
\end{eqnarray}

\begin{eqnarray}
\rho^{\rm I}_{3}(s)&=&-\frac{m_b\langle \bar{q}q\rangle}{64\pi^4}\int_{y_i}^{y_f}dy \int_{z_i}^{1-y}dz \, (y+z)(1-y-z)\left(s-\overline{m}_b^2\right)\left(7s-3\overline{m}_b^2 \right) \, ,
\end{eqnarray}

\begin{eqnarray}
\rho^{\rm I}_{4}(s)&=&-\frac{m_b^2}{2304\pi^4} \langle\frac{\alpha_s GG}{\pi}\rangle\int_{y_i}^{y_f}dy \int_{z_i}^{1-y}dz \left( \frac{z}{y^2}+\frac{y}{z^2}\right)(1-y-z)^3 \left\{ 8s-3\overline{m}_b^2+\overline{m}_b^4\delta\left(s-\overline{m}_b^2\right)\right\} \nonumber\\
&&+\frac{1}{1536\pi^4}\langle\frac{\alpha_s GG}{\pi}\rangle\int_{y_i}^{y_f}dy \int_{z_i}^{1-y}dz (y+z)(1-y-z)^2 \,s\,(5s-4\overline{m}_b^2) \nonumber\\
&&-t\frac{m_b^2}{1152\pi^4}\langle\frac{\alpha_s GG}{\pi}\rangle\int_{y_i}^{y_f}dy \int_{z_i}^{1-y}dz \left(s-\overline{m}_b^2 \right)\left\{ 1-\left( \frac{1}{y}+ \frac{1}{z}\right) (1-y-z) \right. \nonumber\\
&&\left.+ \frac{(1-y-z)^2}{2yz}  -\frac{1-y-z}{2} +\left(\frac{1}{y}+\frac{1}{z} \right)\frac{(1-y-z)^2}{4}
 -\frac{(1-y-z)^3}{12yz}  \right\}\, ,
\end{eqnarray}

\begin{eqnarray}
\rho^{\rm I}_{5}(s)&=&\frac{m_b\langle \bar{q}g_s\sigma Gq\rangle}{128\pi^4}\int_{y_i}^{y_f}dy \int_{z_i}^{1-y}dz   (y+z) \left(5s-3\overline{m}_b^2 \right) \nonumber\\
&&-\frac{m_b\langle \bar{q}g_s\sigma Gq\rangle}{128\pi^4}\int_{y_i}^{y_f}dy \int_{z_i}^{1-y}dz   \left(\frac{y}{z}+\frac{z}{y} \right)(1-y-z) \left(2s-\overline{m}_b^2 \right)  \nonumber\\
&&-t\frac{m_b\langle \bar{q}g_s\sigma Gq\rangle}{1152\pi^4}\int_{y_i}^{y_f}dy \int_{z_i}^{1-y}dz   \left(\frac{y}{z}+\frac{z}{y} \right)(1-y-z) \left(5s-3\overline{m}_b^2 \right) \, ,
\end{eqnarray}

\begin{eqnarray}
\rho^{\rm I}_{6}(s)&=&\frac{m_b^2\langle\bar{q}q\rangle^2}{12\pi^2}\int_{y_i}^{y_f}dy +\frac{g_s^2\langle\bar{q}q\rangle^2}{648\pi^4}\int_{y_i}^{y_f}dy \int_{z_i}^{1-y}dz\, yz \left\{8s-3\overline{m}_b^2 +\overline{m}_b^4\delta\left(s-\overline{m}_b^2 \right)\right\}\nonumber\\
&&-\frac{g_s^2\langle\bar{q}q\rangle^2}{2592\pi^4}\int_{y_i}^{y_f}dy \int_{z_i}^{1-y}dz(1-y-z)\left\{ \left(\frac{z}{y}+\frac{y}{z} \right)3\left(7s-4\overline{m}_b^2 \right)\right.\nonumber\\
&&\left.+\left(\frac{z}{y^2}+\frac{y}{z^2} \right)m_b^2\left[ 7+5\overline{m}_b^2\delta\left(s-\overline{m}_b^2 \right)\right]-(y+z)\left(4s-3\overline{m}_b^2 \right)\right\} \nonumber\\
&&-\frac{g_s^2\langle\bar{q}q\rangle^2}{3888\pi^4}\int_{y_i}^{y_f}dy \int_{z_i}^{1-y}dz(1-y-z)\left\{  \left(\frac{z}{y}+\frac{y}{z} \right)3\left(2s-\overline{m}_b^2 \right)\right. \nonumber\\
&&\left.+\left(\frac{z}{y^2}+\frac{y}{z^2} \right)m_b^2\left[ 1+\overline{m}_b^2\delta\left(s-\overline{m}_b^2\right)\right]+(y+z)2\left[8s-3\overline{m}_b^2 +\overline{m}_b^4\delta\left(s-\overline{m}_b^2\right)\right]\right\}\, ,
\end{eqnarray}

\begin{eqnarray}
\rho^{\rm I}_7(s)&=&\frac{m_b^3\langle\bar{q}q\rangle}{576\pi^2}\langle\frac{\alpha_sGG}{\pi}\rangle\int_{y_i}^{y_f}dy \int_{z_i}^{1-y}dz \left(\frac{y}{z^3}+\frac{z}{y^3}+\frac{1}{y^2}+\frac{1}{z^2}\right)(1-y-z) \nonumber\\
&&\left( 1+\frac{2\overline{m}_b^2}{T^2}\right)\delta\left(s-\overline{m}_b^2\right)\nonumber\\
&&-\frac{m_b\langle\bar{q}q\rangle}{64\pi^2}\langle\frac{\alpha_sGG}{\pi}\rangle\int_{y_i}^{y_f}dy \int_{z_i}^{1-y}dz \left(\frac{y}{z^2}+\frac{z}{y^2}\right)(1-y-z)
\left\{1+\frac{2\overline{m}_b^2}{3}\delta\left(s-\overline{m}_b^2\right) \right\} \nonumber\\
&&-\frac{m_b\langle\bar{q}q\rangle}{192\pi^2}\langle\frac{\alpha_sGG}{\pi}\rangle\int_{y_i}^{y_f}dy \int_{z_i}^{1-y}dz\left\{1+\frac{2\overline{m}_b^2}{3}\delta\left(s-\overline{m}_b^2\right) \right\} \nonumber\\
&&-t\frac{m_b\langle\bar{q}q\rangle}{288\pi^2}\langle\frac{\alpha_sGG}{\pi}\rangle\int_{y_i}^{y_f}dy \int_{z_i}^{1-y}dz\left\{1-\left(\frac{1}{y}+\frac{1}{z}\right)\frac{1-y-z}{2}\right\}\left\{1+\frac{2\overline{m}_b^2}{3}\delta\left(s-\overline{m}_b^2\right) \right\} \nonumber\\
&&-\frac{m_b\langle\bar{q}q\rangle}{384\pi^2}\langle\frac{\alpha_sGG}{\pi}\rangle\int_{y_i}^{y_f}dy \left\{1+\frac{2\widetilde{m}_b^2}{3}\delta\left(s-\widetilde{m}_b^2\right) \right\}  \, ,
\end{eqnarray}

\begin{eqnarray}
\rho^{\rm I}_8(s)&=&-\frac{m_b^2\langle\bar{q}q\rangle\langle\bar{q}g_s\sigma Gq\rangle}{24\pi^2}\int_0^1 dy \left(1+\frac{\widetilde{m}_b^2}{T^2} \right)\delta\left(s-\widetilde{m}_b^2\right)\nonumber\\
&&+\frac{m_b^2\langle\bar{q}q\rangle\langle\bar{q}g_s\sigma Gq\rangle}{96\pi^2}\int_0^1 dy \left( \frac{1}{y}+\frac{1}{1-y} \right)\delta\left(s-\widetilde{m}_b^2\right)\nonumber\\
&&+t\frac{\langle\bar{q}q\rangle\langle\bar{q}g_s\sigma Gq\rangle}{288\pi^2}\int_{y_i}^{y_f} dy \left\{1+\frac{2\widetilde{m}_b^2}{3}\delta\left(s-\widetilde{m}_b^2\right) \right\}  \, ,
\end{eqnarray}

\begin{eqnarray}
\rho^{\rm I}_{10}(s)&=&\frac{m_b^2\langle\bar{q}g_s\sigma Gq\rangle^2}{192\pi^2T^6}\int_0^1 dy \widetilde{m}_b^4\delta \left( s-\widetilde{m}_b^2\right)\nonumber\\
&&-\frac{m_b^4\langle\bar{q}q\rangle^2}{216T^4}\langle\frac{\alpha_sGG}{\pi}\rangle\int_0^1 dy  \left\{ \frac{1}{y^3}+\frac{1}{(1-y)^3}\right\} \delta\left( s-\widetilde{m}_b^2\right)\nonumber\\
&&+\frac{m_b^2\langle\bar{q}q\rangle^2}{72T^2}\langle\frac{\alpha_sGG}{\pi}\rangle\int_0^1 dy  \left\{ \frac{1}{y^2}+\frac{1}{(1-y)^2}\right\} \delta\left( s-\widetilde{m}_b^2\right)\nonumber\\
&&-t\frac{\langle\bar{q}q\rangle^2}{1296}\langle\frac{\alpha_sGG}{\pi}\rangle\int_0^1 dy  \left( 1+\frac{2\widetilde{m}_b^2}{T^2}\right) \delta\left( s-\widetilde{m}_b^2\right)\nonumber\\
&&-\frac{m_b^2\langle\bar{q}g_s\sigma Gq\rangle^2}{384\pi^2T^4}\int_0^1 dy \left( \frac{1}{y}+\frac{1}{1-y}\right)\widetilde{m}_b^2\delta \left( s-\widetilde{m}_b^2\right)\nonumber\\
&&-t\frac{\langle\bar{q}g_s\sigma Gq\rangle^2}{1728\pi^2}\int_0^1 dy \left(1+\frac{3\widetilde{m}_b^2}{2T^2}+\frac{\widetilde{m}_b^4}{T^4} \right)\delta \left( s-\widetilde{m}_b^2\right)\nonumber\\
&&-t\frac{\langle\bar{q}g_s\sigma Gq\rangle^2}{2304\pi^2}\int_0^1 dy \left(1+\frac{2\widetilde{m}_b^2}{T^2}  \right)\delta \left( s-\widetilde{m}_b^2\right)\nonumber\\
&&+\frac{m_b^2\langle\bar{q}q\rangle^2}{216T^6}\langle\frac{\alpha_sGG}{\pi}\rangle\int_0^1 dy  \widetilde{m}_b^4  \delta\left( s-\widetilde{m}_b^2\right) \, ,
\end{eqnarray}

\begin{eqnarray}
\rho^{\rm II}_{0}(s)&=&\frac{1}{3072\pi^6s}\int_{y_i}^{y_f}dy \int_{z_i}^{1-y}dz \, yz\, (1-y-z)^3\left(s-\overline{m}_b^2\right)^2\left(49s^2-30s\overline{m}_b^2+\overline{m}_b^4 \right)  \nonumber\\
&&+\frac{1}{3072\pi^6s} \int_{y_i}^{y_f}dy \int_{z_i}^{1-y}dz \, yz \,(1-y-z)^2\left(s-\overline{m}_b^2\right)^3\left(3s+\overline{m}_b^2\right)    \, ,
\end{eqnarray}

\begin{eqnarray}
\rho_{3}^{\rm II}(s)&=&-\frac{m_b\langle \bar{q}q\rangle}{16\pi^4}\int_{y_i}^{y_f}dy \int_{z_i}^{1-y}dz \, (y+z)(1-y-z)\left(s-\overline{m}_b^2\right)  \, ,
\end{eqnarray}

\begin{eqnarray}
\rho_{4}^{\rm II}(s)&=&-\frac{m_b^2}{2304\pi^4s} \langle\frac{\alpha_s GG}{\pi}\rangle\int_{y_i}^{y_f}dy \int_{z_i}^{1-y}dz \left( \frac{z}{y^2}+\frac{y}{z^2}\right)(1-y-z)^3 \nonumber\\
&&\left\{ 8s-\overline{m}_b^2+\frac{5\overline{m}_b^4}{3}\delta\left(s-\overline{m}_b^2\right)\right\} \nonumber\\
&&-\frac{m_b^2}{2304\pi^4s}\langle\frac{\alpha_s GG}{\pi}\rangle\int_{y_i}^{y_f}dy \int_{z_i}^{1-y}dz \left(\frac{z}{y^2}+\frac{y}{z^2} \right) (1-y-z)^2 \, \overline{m}_b^2 \nonumber\\
&&-\frac{1}{9216\pi^4s} \langle\frac{\alpha_s GG}{\pi}\rangle\int_{y_i}^{y_f}dy \int_{z_i}^{1-y}dz \left( y+z\right)(1-y-z)^2 \left( 5s^2-3\overline{m}_b^4\right) \nonumber\\
&&+\frac{1}{4608\pi^4s} \langle\frac{\alpha_s GG}{\pi}\rangle\int_{y_i}^{y_f}dy \int_{z_i}^{1-y}dz \left( y+z\right)(1-y-z) \left( s^2-\overline{m}_b^4\right) \nonumber\\
&&+\frac{1}{2304\pi^4} \langle\frac{\alpha_s GG}{\pi}\rangle\int_{y_i}^{y_f}dy \int_{z_i}^{1-y}dz \left( y+z\right)(1-y-z)^2 \left( 5s-4\overline{m}_b^2\right) \nonumber\\
&&+\frac{1}{41472\pi^4s} \langle\frac{\alpha_s GG}{\pi}\rangle\int_{y_i}^{y_f}dy \int_{z_i}^{1-y}dz \, (1-y-z)^3 \left( 55s^2-48s\overline{m}_b^2+3\overline{m}_b^4\right)  \nonumber\\
&&+\frac{1}{6912\pi^4s} \langle\frac{\alpha_s GG}{\pi}\rangle\int_{y_i}^{y_f}dy \int_{z_i}^{1-y}dz  \, yz\,(1-y-z) \left( 5s^2-3\overline{m}_b^4\right) \nonumber\\
&&-\frac{1}{3456\pi^4s} \langle\frac{\alpha_s GG}{\pi}\rangle\int_{y_i}^{y_f}dy \int_{z_i}^{1-y}dz \, (1-y-z)^2 \left( s-\overline{m}_b^2\right) \left( 2s-\overline{m}_b^2\right)  \nonumber\\
&&+\frac{1}{1728\pi^4s} \langle\frac{\alpha_s GG}{\pi}\rangle\int_{y_i}^{y_f}dy \int_{z_i}^{1-y}dz \, yz \left( s-\overline{m}_b^2\right)\left( 2s-\overline{m}_b^2\right) \, ,
\end{eqnarray}

\begin{eqnarray}
\rho^{\rm II}_{5}(s)&=&\frac{m_b\langle \bar{q}g_s\sigma Gq\rangle}{64\pi^4}\int_{y_i}^{y_f}dy \int_{z_i}^{1-y}dz  \, (y+z)\nonumber\\
&&
-\frac{m_b\langle \bar{q}g_s\sigma Gq\rangle}{288\pi^4}\int_{y_i}^{y_f}dy \int_{z_i}^{1-y}dz  \,  (1-y-z)      \, ,
\end{eqnarray}

\begin{eqnarray}
\rho_{6}^{\rm II}(s)&=& \frac{g_s^2\langle\bar{q}q\rangle^2}{648\pi^4s}\int_{y_i}^{y_f}dy \int_{z_i}^{1-y}dz\, yz \left\{8s-\overline{m}_b^2 +\frac{5\overline{m}_b^4}{3}\delta\left(s-\overline{m}_b^2 \right)\right\}\nonumber\\
&&+\frac{g_s^2\langle\bar{q}q\rangle^2}{1944\pi^4s}\int_{y_i}^{y_f}dy \,y(1-y) \,\widetilde{m}_b^2   \nonumber\\
&&-\frac{g_s^2\langle\bar{q}q\rangle^2}{1296\pi^4}\int_{y_i}^{y_f}dy \int_{z_i}^{1-y}dz \, (1-y-z)\left\{ 3\left(\frac{z}{y}+\frac{y}{z} \right)  +\left(\frac{z}{y^2}+\frac{y}{z^2} \right)m_b^2  \delta\left(s-\overline{m}_b^2 \right)\right.\nonumber\\
&&\left. +(y+z)\left[8+2\overline{m}_b^2\delta\left(s-\overline{m}_b^2\right) \right] \right\} \nonumber\\
&&-\frac{g_s^2\langle\bar{q}q\rangle^2}{11664\pi^4s}\int_{y_i}^{y_f}dy \int_{z_i}^{1-y}dz \, (1-y-z)\left\{  27\left(\frac{z}{y}+\frac{y}{z} \right)s+11\left(\frac{z}{y^2}+\frac{y}{z^2} \right)\right. \nonumber\\
&&\left.m_b^2\overline{m}_b^2\delta\left(s-\overline{m}_b^2\right)+(y+z)\left[6\left(8s-\overline{m}_b^2\right) +10\overline{m}_b^4\delta\left(s-\overline{m}_b^2\right)\right] \right\}\, ,
\end{eqnarray}

\begin{eqnarray}
\rho_7^{\rm II}(s)&=&\frac{m_b^3\langle\bar{q}q\rangle}{288\pi^2 T^2}\langle\frac{\alpha_sGG}{\pi}\rangle\int_{y_i}^{y_f}dy \int_{z_i}^{1-y}dz \left(\frac{y}{z^3}+\frac{z}{y^3}+\frac{1}{y^2}+\frac{1}{z^2}\right)(1-y-z) \delta\left(s-\overline{m}_b^2\right)\nonumber\\
&&-\frac{m_b\langle\bar{q}q\rangle}{96\pi^2}\langle\frac{\alpha_sGG}{\pi}\rangle\int_{y_i}^{y_f}dy \int_{z_i}^{1-y}dz \left(\frac{y}{z^2}+\frac{z}{y^2}\right)(1-y-z)  \delta\left(s-\overline{m}_b^2\right)\nonumber\\
&&-\frac{m_b\langle\bar{q}q\rangle}{288\pi^2}\langle\frac{\alpha_sGG}{\pi}\rangle\int_{y_i}^{y_f}dy \int_{z_i}^{1-y}dz\delta\left(s-\overline{m}_b^2\right) \nonumber\\
&&-\frac{m_b\langle\bar{q}q\rangle}{864\pi^2}\langle\frac{\alpha_sGG}{\pi}\rangle\int_{y_i}^{y_f}dy \int_{z_i}^{1-y}dz\left(\frac{1-y}{y}+\frac{1-z}{z}\right)
\delta\left(s-\overline{m}_b^2\right) \nonumber \\
&&-\frac{m_b\langle\bar{q}q\rangle}{576\pi^2}\langle\frac{\alpha_sGG}{\pi}\rangle\int_{y_i}^{y_f}dy   \delta \left(s-\widetilde{m}_b^2\right) \, ,
\end{eqnarray}
where the superscripts  I and  II   denote the $C\gamma_5-C\gamma_\mu$ type and $C\gamma_\mu-C\gamma_\nu$ type tetraquark states, respectively; $y_{f}=\frac{1+\sqrt{1-4m_b^2/s}}{2}$,
$y_{i}=\frac{1-\sqrt{1-4m_b^2/s}}{2}$, $z_{i}=\frac{y
m_b^2}{y s -m_b^2}$, $\overline{m}_b^2=\frac{(y+z)m_b^2}{yz}$,
$ \widetilde{m}_b^2=\frac{m_b^2}{y(1-y)}$, $\int_{y_i}^{y_f}dy \to \int_{0}^{1}dy$, $\int_{z_i}^{1-y}dz \to \int_{0}^{1-y}dz$ when the $\delta$ functions $\delta\left(s-\overline{m}_b^2\right)$ and $\delta\left(s-\widetilde{m}_b^2\right)$ appear.
The condensates $\langle \frac{\alpha_s}{\pi}GG\rangle$, $\langle \bar{q}q\rangle\langle \frac{\alpha_s}{\pi}GG\rangle$,
$\langle \bar{q}q\rangle^2\langle \frac{\alpha_s}{\pi}GG\rangle$, $\langle \bar{q} g_s \sigma Gq\rangle^2$ and $g_s^2\langle \bar{q}q\rangle^2$ are the vacuum expectations
of the operators of the order
$\mathcal{O}(\alpha_s)$.  The four-quark condensate $g_s^2\langle \bar{q}q\rangle^2$ comes from the terms
$\langle \bar{q}\gamma_\mu t^a q g_s D_\eta G^a_{\lambda\tau}\rangle$, $\langle\bar{q}_jD^{\dagger}_{\mu}D^{\dagger}_{\nu}D^{\dagger}_{\alpha}q_i\rangle$  and
$\langle\bar{q}_jD_{\mu}D_{\nu}D_{\alpha}q_i\rangle$, rather than comes from the perturbative corrections of $\langle \bar{q}q\rangle^2$.
 The condensates $\langle g_s^3 GGG\rangle$, $\langle \frac{\alpha_s GG}{\pi}\rangle^2$,
 $\langle \frac{\alpha_s GG}{\pi}\rangle\langle \bar{q} g_s \sigma Gq\rangle$ have the dimensions 6, 8, 9 respectively,  but they are   the vacuum expectations
of the operators of the order    $\mathcal{O}( \alpha_s^{3/2})$, $\mathcal{O}(\alpha_s^2)$, $\mathcal{O}( \alpha_s^{3/2})$ respectively, and discarded.  We take
the truncations $n\leq 10$ and $k\leq 1$ in a consistent way,
the operators of the orders $\mathcal{O}( \alpha_s^{k})$ with $k> 1$ are  discarded. Furthermore,  the values of the  condensates $\langle g_s^3 GGG\rangle$, $\langle \frac{\alpha_s GG}{\pi}\rangle^2$,
 $\langle \frac{\alpha_s GG}{\pi}\rangle\langle \bar{q} g_s \sigma Gq\rangle$   are very small, and they can be  neglected safely.


\begin{thebibliography}{99}

\bibitem{Belle1105} I. Adachi et al,  arXiv:1105.4583.

\bibitem{Belle1110}   A. Bondar   et al,  Phys. Rev. Lett. {\bf 108} (2012) 122001.


\bibitem{Belle1308} P. Krokovny  et al, Phys. Rev. {\bf D88} (2013)  052016.

\bibitem{Molecule-Zb} A. E. Bondar, A. Garmash, A. I. Milstein, R. Mizuk and M. B. Voloshin, Phys. Rev. {\bf D84} (2011) 054010;
J. R. Zhang, M. Zhong and M. Q. Huang,  Phys. Lett. {\bf B704} (2011) 312;
M. B. Voloshin,  Phys. Rev. {\bf D84} (2011) 031502;
J. Nieves and M. Pavon Valderrama, Phys. Rev. {\bf D84} (2011) 056015;
Z. F. Sun, J. He, X. Liu, Z. G. Luo and S. L. Zhu, Phys. Rev. {\bf D84} (2011) 054002;
M. Cleven, F. K. Guo, C. Hanhart and Ulf-G. Meissner, Eur. Phys. J. {\bf A47} (2011) 120;
 T. Mehen and J. W. Powell, Phys. Rev. {\bf D84} (2011) 114013;
Y. Yang , J. Ping, C. Deng and H. S. Zong, J. Phys. {\bf G39} (2012) 105001;
S. Ohkoda, Y. Yamaguchi, S. Yasui, K. Sudoh and A. Hosaka, Phys. Rev. {\bf D86} (2012) 014004;
 H. W. Ke, X. Q. Li, Y. L. Shi, G. L. Wang and X. H. Yuan, JHEP {\bf 1204} (2012) 056;
Y. Dong, A. Faessler, T. Gutsche and V. E. Lyubovitskij, J. Phys. {\bf G40} (2013) 015002;
 M. B. Voloshin,  Phys. Rev. {\bf D87} (2013) 074011.


\bibitem{Tetraquark-Zb} A. Ali and C. Hambrock and W. Wang,  Phys. Rev. {\bf D85} (2012) 054011.


\bibitem{Tetraquark-Zb-QCDSR} C. Y. Cui, Y. L. Liu and M. Q. Huang, Phys. Rev. {\bf D85} (2012) 074014.

\bibitem{Cusp-Zb} D. V. Bugg, Europhys. Lett. {\bf 96} (2011) 11002.


\bibitem{Rescatter-Zb}  D. Y. Chen, X. Liu and S. L. Zhu, Phys. Rev. {\bf D84} (2011) 074016;
G. Li, F. l. Shao, C. W. Zhao and Q. Zhao,  Phys. Rev. {\bf D87} (2013) 034020.



\bibitem{BES3900}   M. Ablikim et al, Phys. Rev. Lett. {\bf 110} (2013) 252001.

\bibitem{Belle3900} Z. Q. Liu   et al, Phys. Rev. Lett. {\bf 110} (2013) 252002.

\bibitem{CLEO3900} T. Xiao, S. Dobbs, A. Tomaradze and K. K. Seth,  Phys. Lett. {\bf B727} (2013) 366.



\bibitem{BES1308}   M. Ablikim  et al, Phys. Rev. Lett. {\bf 112} (2014) 132001.

\bibitem{BES1309}  M. Ablikim  et al, Phys. Rev. Lett. {\bf 111} (2013) 242001.


 \bibitem{WangHuangTao} Z. G. Wang and T. Huang,  Phys. Rev. {\bf D89} (2014) 054019.

 \bibitem{Wang1311} Z. G. Wang,  Eur. Phys. J. {\bf C74} (2014) 2874.

 \bibitem{Wang1312} Z. G. Wang,  arXiv:1312.1537.


 \bibitem{Wang-Axial} Z. G. Wang, Eur. Phys. J. {\bf C70} (2010) 139.

 \bibitem{Chenwei} W. Chen  and S. L. Zhu, Phys. Rev. {\bf D83} (2011) 034010.

\bibitem{SVZ79}  M. A. Shifman, A. I. Vainshtein and V. I. Zakharov, Nucl. Phys. {\bf B147} (1979) 385.

\bibitem{Reinders85} L. J. Reinders, H. Rubinstein and S. Yazaki, Phys. Rept. {\bf 127} (1985) 1.

\bibitem{WangHcHb} Z. G. Wang, Eur. Phys. J. {\bf C73} (2013) 2533.

\bibitem{Ioffe2005} B. L. Ioffe, Prog. Part. Nucl. Phys. {\bf 56} (2006) 232.

\bibitem{ColangeloReview} P. Colangelo and A. Khodjamirian, hep-ph/0010175.

\bibitem{PDG}   J. Beringer et al, Phys. Rev. {\bf D86} (2012) 010001.

\bibitem{Wang1301} Z. G. Wang,  JHEP {\bf 1310} (2013) 208.


\bibitem{Maiani-2014}  L. Maiani, F. Piccinini, A. D. Polosa and V. Riquer, Phys. Rev. {\bf D89} (2014) 114010;
 M. Nielsen and F. S. Navarra,  Mod. Phys. Lett. {\bf  A29} (2014) 1430005;
  Z. G. Wang,  arXiv:1405.3581.

 \bibitem{Wang4140} Z. G. Wang, Eur. Phys. J. {\bf C63} (2009) 115;
Z. G. Wang, Z. C. Liu and X. H. Zhang, Eur. Phys. J. {\bf C64} (2009) 373;
Z. G. Wang and X. H. Zhang,  Commun. Theor. Phys. {\bf 54} (2010) 323;
Z. G. Wang and X. H. Zhang, Eur. Phys. J. {\bf C66} (2010) 419.

\bibitem{Nielsen3900} J. M. Dias, F. S. Navarra, M. Nielsen and C. M. Zanetti, Phys. Rev. {\bf D88} (2013) 016004.


\bibitem{RCVerma} R. C. Verma,  J. Phys. {\bf G39} (2012) 025005.




\end{thebibliography}
\end{document}